\documentclass[prb,onecolumn,nofootinbib,citeautoscript,10pt,longbibliography,notitlepage]{revtex4-2}
\usepackage{graphicx}
\usepackage{dcolumn}
\usepackage{bm,amsfonts,amsmath}
\usepackage{color} 
\usepackage[tight]{subfigure}

\usepackage[papersize={8.5in,11in}]{geometry}
\usepackage{xcolor}
\usepackage{tabularx}

\definecolor{darkblue}{rgb}{0.,0.,0.4}
\definecolor{darkred}{rgb}{0.5,0.,0.}
\definecolor{BlueViolet}{RGB}{138,43,226}
\definecolor{SkyBlue}{RGB}{30,144,255}
\definecolor{DarkGreen}{RGB}{0,100,0}
\usepackage[pdftex,colorlinks=true,linkcolor=darkblue,citecolor=blue,urlcolor=darkred]{hyperref}
\usepackage{float}
\usepackage{physics}
\usepackage{cancel}
\usepackage[version=4]{mhchem}

\newcommand{\bp}{\mathbf{p}}
\newcommand{\bq}{\mathbf{q}}
\newcommand{\br}{\mathbf{r}}

\def \nn{\nonumber \\}

\begin{document}

\title{Generic deformation channels for critical Fermi surfaces in the collisionless regime}

\author{Kazi Ranjibul Islam}
\affiliation{School of Physics and Astronomy and William I. Fine Theoretical Physics Institute,
University of Minnesota, Minneapolis, MN 55455, USA}

\author{Ipsita Mandal}
\affiliation{Institute of Nuclear Physics, Polish Academy of Sciences, 31-342 Krak\'{o}w, Poland}

\begin{abstract}
Using a quantum Boltzmann equation framework, we analyse the nature of generic low-energy deformations of a critical Fermi surface, which exists at the non-Fermi liquid fixed point of a system consisting of fermions interacting with massless bosons. The non-Fermi liquid behaviour arises due to the itinerant quasiparticles of the Fermi surface interacting strongly with the massless bosons, which on the other hand undergo Landau damping as a result of the mutual interactions. Focussing on the collisionless regime, where we neglect the collision integral, we chalk out the possible excitations spanning the entire spectrum of angular momentum ($\ell$) channels (i.e., including both small and large values of $\ell$). The excitations are of two types: particle-hole like localized excitations forming an energy band (or continuum) and delocalized collective modes with discrete energies. Although we find a collective mode analogous to the zero sound of a Fermi liquid, its dispersion shows a crossover from a $\Omega \sim |\mathbf q|^{6/5}$ behaviour to the usual $ \Omega \sim |\mathbf q|$ dependence, where $\Omega$ and $\mathbf q$ represent the frequency and momentum, respectively. We estimate the frequency scale at which this crossover takes place. We also determine the boundary for the particle-hole continuum in the $\Omega$--$\mathbf q$ plane, and observe a crossover from $\Omega \sim |\mathbf q|^{3/2}$ to $ \Omega \sim |\mathbf q|$ behaviour, determined by another crossover frequency scale.
\end{abstract}
\maketitle

\tableofcontents


 \section{Introduction}
 
 One of the long standing puzzles of condensed matter research is to understand non-Fermi liquid (NFL) metals \cite{metlsach1,metlsach,olav,Lee-Dalid,ips-uv-ir1,ips-uv-ir2,ips-fflo,ips-nfl-u1,ips-rafael}, which goes beyond the framework of the conventional Landau's Fermi liquid theory. The absence of well-defined low-energy quasiparticles makes both the thermodynamics and transport properties \cite{ips-subir,ips-c2,ips-hermann,ips-hermann2,ips-hermann3} of these metals quite different, compared to the normal metals. For example, instead of a conventional quadratic dependence on the temprature $T$, the resistivity shows a linear-in-$T$ scaling. Other manifestations of an NFL behaviour involve changed scalings for optical conductivity \cite{ips-subir} and enhanced susceptibility towards superconducting instability \cite{ips2,ips-sc,ips-c2}. NFL character can emerge in various scenarios, for instance in (1) finite-density fermions interacting with a massless boson arising at a quantum critical point \cite{metlsach1,metlsach,Lee-Dalid,ips-uv-ir1,ips-uv-ir2,ips-fflo,ips-rafael}, or with massless gauge field(s) \cite{olav,ips2,ips3,ips-nfl-u1}; (2) Fermi level tuned at the band-touching points of semimetals in the presence of long-ranged Coulomb interactions \cite{Abrikosov,moon-xu,rahul-sid,ips-rahul,malcolm-bitan,ips-birefringent}. In this paper, we will focus on the NFL arising at the two-dimensional (2d) Ising-nematic quantum critical point (QCP) \cite{metlsach1,metlsach,Lee-Dalid,ips-uv-ir1,ips-uv-ir2}.

Ising-nematic ordering implies the spontaneous breaking of the rotational symmetry of a Fermi surface in the $xy$-plane, which describes the Pomeranchuk instability in the $\ell=2$ angular momentum channel. In other words, the four-fold rotational symmetry of the Fermi surface (i.e., the symmetry for rotations by $\pi/2$) is broken down to two-fold rotations, such that the $x$- and $y$-directions become anisotropic \cite{metlsach1,sachdev_2011}. This broken symmetry is explained by invoking the Ising-nematic order parameter, which can be captured by a real scalar boson $\phi$, centred at wavevector $\mathbf{Q} = 0$. Right at the quantum phase transition, the fermionic system has a well-defined Fermi surface, but no well-defined quasiparticles \cite{Lee-Dalid,ips-uv-ir1,ips-uv-ir2,ips-subir}. The Ising-nematic phase is believed to be present in various strongly-correlated systems, like cuprate superconductors \cite{yoichi,hinkov,Kohsaka,Daou} and Fe-based superconductors \cite{hsu2008superconductivity, margadonna2008crystal,mcqueen2009tetragonal}. Hence, an in-depth understanding of the origins and behaviour of this quantum critical point is quintessential.

Even though Landau quasiparticles are absent in the Ising-nematic QCP scenario, it is interesting to ask what kind of excitations of the critical Fermi surface may exist. This comes under the broad topic of whether NFL metal supports collective modes like zero sound obtained in the Fermi liquid case. Recent studies \cite{ips-zero-mode,debanjan,else} have investigated this aspect using various theoretical and numerical techniques. In this paper, we continue the study \cite{ips-zero-mode} initiated by one of us. The problem is analysed by decomposing the generic Fermi surface displacements in angular momentum channels (labelled by $\ell$), as angular momnetum is a good quantum number in two spatial dimensions. Ref.~\cite{ips-zero-mode} computed the dispersion of the collective modes originating from angular momentum channels below a critical value $\ell_{crit}$. Here, we compute generic Fermi surface displacements which might belong to collective excitaions or particle-hole like continuum, and are not restricted to $\ell <\ell_{crit}$.  

In a Fermi liquid, a standard technique used to derive the displacements of the Fermi surface is the quantum Boltzmann equation (QBE) formalism, which hinges on the existence of well-defined quasiparticles. However, since quasiparticles get destroyed in an NFL, a naive application of this framework is inadmissible. To overcome this difficulty, we use the nonequilibrium Green's function technique outlined in Refs.~\cite{prange,kim_qbe,ips-zero-mode}. This formalism introduces a generalized Landau-interaction, which has a frequency-dependence in addition to the usual angular-dependence (seen in a Fermi liquid). The nontrivial dependence on frequency is due to the Landau-damped bosonic propagator, which mediates strong interactions in the fermionic system. In this paper, we supplement the calculations and results of Ref.~\cite{ips-zero-mode} by not confining ourselves to the low angular momentum channels. This automatically comes about as we implement the full form (i.e., without using an approximation for its integral form) of the generalized Landau parameter for the NFL.

The paper is organized as follows. In Sec.~\ref{secmodel}, we introduce the model for the Ising-nematic quantum critical point, which shows an NFL fixed point \cite{Lee-Dalid,ips-uv-ir1,ips-uv-ir2,ips-subir,ips-sc,ips-zero-mode}. We also review the form of the Green's functions in the Keldysh formalism, which is useful for setting up the QBE for the fermions. In Sec.~\ref{secqbe}, we demonstrate the derivation of the QBE, and write down the final form of the resulting recursive equations. In order to find simple analytical expressions for the Fermi surface displacements, we consider a simple model built with the help of some reasonable assumptions. This enables us to find the Fermi surface displacements analytically in Sec.~\ref{secsoln}. Sec.~\ref{secnum} deals with incorporating a more complete computational framework, without the simplifying truncations made in Sec.~\ref{secsoln}. In this case, the solutions are found numeircally, as closed form analytical expressions are not possible. Finally, we end with a summary and outlook in Sec.~\ref{secsum}.

\section{Model}
\label{secmodel}

In this section, we review the details necessary for setting up the QBE for the Ising-nematic quantum critical point, which was studied in Ref.~\cite{ips-zero-mode} in the strictly collisionless limit.
The Ginzburg-Landau action for the real boson order parameter is given by \cite{metlsach1,sachdev_2011}
\begin{align}
\label{ac1}
S_\phi & =  \frac{1}{2} \int d\tau \,dx \, dy\left[  \phi(\tau,x,y)
\left(- \partial_\tau ^2 -  c_\phi ^2 \, \partial_x ^2 -  c_\phi ^2 \, \partial_y ^2
\right) \phi (\tau,x,y)
 + r_\phi \,\phi^2 (\tau,x,y)
+\frac{u_\phi \,\phi^ 4(\tau,x,y) } {12}
\right],
\end{align}
in the position space [spanned by the $(x,\,y)$ coordinates] and in imaginary time $\tau$. Here,
$c_\phi$ is the boson velocity, $ r_\phi$ is the parameter tuning across the phase transition, and $u_\phi$ is the coupling constant for the $\phi^4$-term.
One can show that all couplings can be scaled away or set equal to unity, except $r_\phi$. 
The quantum phase transition is the zero-temperature phase transition appearing at $ r_\phi =0$, and the nomenclature ``Ising-nematic'' stems from the fact that the above action is in the same universality class as the classical 2d Ising model. The physics of this purely bosonic part, however, is not the complete story, as the fermions coupling to the gapless bosons at $ r_\phi =0$ change the nature of the quantum critical fluctuations. The resulting composite system is strongly coupled, with the bosons acquiring Landau damping, causing the fermions to behave as an NFL in turn. Although the NFL fixed point emerges at temperature $T=0$, which of course is not observable, the quantum effects show their distinctive NFL-like features  in an extended fan-shaped quantum critical region emanating from the QCP \cite{sachdev_2011}.

For the purely fermionic part, we use the patch theory introduced in Refs.~\cite{metlsach1,sachdev_2011,Lee-Dalid,ips-uv-ir1,ips-uv-ir2,ips-nfl-u1}, which turns out to be very useful to incorporate the fact that fermions around a point of the Fermi surface primarily couple with the bosons with momentum $\mathbf q$ tangential to the Fermi surface. In other words, fermions in different momentum patches (except for
the ones at the antipodal points) are decoupled from each other in the low-energy limit, as they have non-parallel tangent vectors. This implies that the properties of local observables (e.g., Green’s functions) can be extracted from local patches in the momentum space, and the global properties of the Fermi surface have no effect.
In this coordinate system, the action for the fermions is captured by
\begin{align}
\label{ac2}
S_\psi &=  
\int_{k} \psi^\dagger (k) 
\left ( -i \,k_0 + \delta_k \right ) \psi (k) \,,
\quad \delta_k =  v_F \,k_{\perp} +  \frac{k_{\parallel}^2} {2\,m}\,,
\end{align}
in the Fourier space, where $\psi$ denotes the fermionic field residing on the patch under consideration. We have used the condensed notations $k$ to indicate $ (k_0, \mathbf k)$ (with $k_0 $ denoting the Matsubara frequency) and $ \int_k $ to indicate the integral $ \int \frac{dk_0\,dk_{\parallel}\,dk_\perp } {(2\,\pi)^3}$.
While writing the kinetic terms in the the patch coordinates, we have expanded the fermion momentum about the local Fermi momentum $k_F$, such that $ k_{\perp}$ is directed perpendicular to the local Fermi surface, and $k_{\parallel}$ is tangential to it.

The last step involves writing down the coupling term between the bosons and the fermions. A simple convenient choice, dictated by symmetry considerations, is captured by the action \cite{sachdev_2011}
\begin{align}
\label{ac3}
 S_{int} = \tilde e
\int_{k} \int_q \left(  \cos k_x -\cos k_y  \right)
\phi(q) \,\psi^\dagger (k+q) \, \psi(k) \,,
\end{align}
where $\tilde e$ is the fermion-boson coupling constant. We would like to point out that this piece of the total action is written in the global momentum space coordinates (rather than the patch coordinates of the Fermi surface). The form factor $\left(  \cos k_x -\cos k_y  \right)$ has a d-wave sysmmetry, reflecting the fact that that the bosonic order parameter causes a quadrupolar distortion of the Fermi surface. Since the low-energy effective action involves only small momentum transfers, the integral over $q$ is over small momenta. On the other hands, in the global coordinate system, $k$ extends over the entire Brillouin zone.
Converting the above to the patch coordinates, and keeping only the leading order terms in momentum about the Fermi surface, we get
$\left(  \cos k_x -\cos k_y  \right) \simeq \cos k_F$. Re-expressing the coupling strength as $e = \tilde e \, \cos k_F$, we then get the following form for the fermion-boson interaction in the patch coordinates:
\begin{align}
 \tilde S_{int} 
= e\,  \int_{k}  \int_q \phi(q) \,\psi^\dagger (k+q) \, \psi(k) \,.
\end{align}

The final form of the total action $S_{tot}$ is obtained after an appropriate rescaling of the energy and momenta, and dropping the irrelevant terms \cite{Lee-Dalid,ips-uv-ir1} after determining the engineering dimension of each term. This is done by setting the dimensions $[k_0] = [k_\perp] = 1$ and $ [k_\parallel] =1/2$. This gives us
\begin{align}
S_{tot} & = S_\psi + \tilde S_\phi + \tilde S_{int} \,,\quad
\tilde S_\phi  = \frac{1}{2} \int_{k}\,k_{\parallel}^2 \,\phi(-k) \,\phi(k) \,.
\end{align}
The reason for invoking the patch coordinates is that it must be used to derive the correct form of the one-loop self-energies of the bosons and fermions \cite{metlsach1,Lee-Dalid,ips-uv-ir1}.

Using $S_{tot}$, we find that the bare Matusubara Green's functions for the fermions and bosons take the forms:
\begin{align}
\label{eqbareGF}
\mathcal{G}^{(0)} (k)
= \dfrac{1} { i\, k_0 - \delta_k  } \text{ and }
\mathcal{D}^{(0)} (q) = \dfrac{1} { -q_{||}^2} \,,
\end{align}
respectively. Since the bare kinetic term of the boson depends only on $q_{||}$, we
need to include the lowest-order quantum corrections to ensure that the loop-integrals are infrared- and ultraviolet-finite \cite{Lee-Dalid,ips-uv-ir1}. Hence, we use the one-loop corrected bosonic propagator
 \begin{align}
  \mathcal{D}^{(1)}(q)=\left[
 \left(  \mathcal{D}^{(0)} (q)\right) ^{-1} 
-\Pi(q)   \right]^{-1}
 =-\dfrac{1} {q_{||}^2 + \dfrac{e^2  \, m\, |q_0|} {2 \,\pi \,v_F \,|q_{||}|}} \,,
\label{eqbareD}
\end{align} 
which includes the one-loop bosonic self-energy
$ \Pi(q)= - e^2 \int_k \mathcal{G} ^{(0)}(k+q)
 \,\mathcal{G} ^{(0)} (k)=\dfrac{e^2\, m \,|q_0|}
 {2 \, \pi \, v_F \,|q_{||}|}$,
that also accounts for the Landau damping. We note that this expression is valid in the limits $| q_0|/|q_\parallel| \ll 1 $, $|\mathbf q| \ll k_F$, and $ |\mathbf q| \rightarrow 0 $ (see, for example, below Eqs.~(26) and (27) of Ref.~\cite{ips-rafael}).

Plugging in the dressed bosonic propagator, the one-loop fermion self-energy in the Matsubara space turns out to be 
\begin{align}
\Sigma(k)=- \dfrac{  i\,   e^{4/3}\, \text{sgn}(k_0) \, |k_0|^{2/3}}
{2 \, \sqrt{3}\, \pi^{2/3}\,k_F^{1/2}} \,.
\label{selfenergyform}
\end{align}
The imaginary part of the self-energy represents the lifetime of the underlying quasiparticles. For a Fermi liquid, the self-energy turns out to be proportional to $ \omega^2$, which is much smaller than the excitation energy $\omega$, in the limit $\omega \ll 1$. As a result, the quasiparticles are well-defined for a Fermi liquid. However, here we find that $\text{Im}\Sigma \propto |\omega|^{2/3}$, which is greater than $|\omega|$ in the $\omega \rightarrow 0$ limit, which is why the quasiparticle-picture does not hold for the Ising-nematic quantum critical point. As a result, the traditional use of Boltzmann equation to derive the kinetic equation is not valid.

To deal with a generic nonquilibrium situation, a widely used method is to use the closed-time Keldysh contour \cite{kamenev}. Denoting the field operators on the forward and the backward branches of the time contour by the subscripts ``$+$'' and ``$-$'', respectively, the total action is given by $
S_{K, tot} = S_{K,\psi} + S_{K,\phi} +S_{K,int}$, where
\begin{align}
\label{kelaction}
   S_{K,\psi} &=\int_{-\infty}^{\infty} dt
   \int \dfrac{d^2\mathbf k}   {( 2\, \pi  )^2}
  \left[ {\psi}^\dagger_+ (t,\mathbf k)
  \left(i \,\partial_t-\delta_k \right)\psi_+ (t,\mathbf k)
 -{\psi}^\dagger_- (t,\mathbf k) \left(i \,\partial_t-\delta_k\right)
 \psi_- (t,\mathbf k) \right], \nn
S_{K,\phi} & =\dfrac{1}{2} \int_{-\infty}^{\infty} dt
 \int \dfrac{d^2 \mathbf k} {( 2\, \pi  )^2}
 \left[\phi_+(t,-\mathbf k)
 \left(- \partial_t^2-k_\parallel^2\right)\phi_+(t,\mathbf k)
 -\phi_-(t,-\mathbf k)\left(- \partial_t^2- k_\parallel^2\right)\phi_-(t,\mathbf k)\right], \nn
S_{K,int} & =-e
\int_{-\infty}^{\infty} dt
\int d^2\br
\left[\phi_+(t,\br)\,{\psi}^\dagger_+(t,\br) \,\psi_+(t,\br)
-\phi_-(t,\br) \,{\psi}^\dagger_-(t,\br) \, \psi_-(t,\br)\right].
\end{align}
The relative minus sign in each part comes from reversing the direction of the time integration on the backward branch of the contour. 

The Keldysh formalism automatically includes different Green’s functions, defined according to the location of the time argument on the contour, which we show in Appendex~\ref{app1} for the sake of completeness. 
As a final step, we change the coordinate system to $(t_{rel}=t_1-t_2, \, \mathbf r_{rel} =\mathbf r_1 -\mathbf r_2)$ and $(t=\frac{t_1+t_2} {2}, \,  \mathbf r  = \frac{ \mathbf r_1+ \mathbf r_2} {2})$ (also referred to as the Wigner coordinates), as it becomes convenient to go to the equilibrium limit, where the dependence on $(t, \mathbf r)$ drops out. In other words, all Green's functions of a translationally invariant system (including both time and space translations) depend only on the differences $t_{rel} =t_1-t_2$ and $\mathbf{r}_{rel} =\mathbf r_1-\mathbf r_2$. Let $k_1\equiv (\omega_{k_1}, \, \mathbf k_1)$ and $k_2\equiv (\omega_{k_2}, \, \mathbf k_2)$ be the energy-momentum variables conjugate to $x_1\equiv(t_1,\mathbf r_1)$ and $x_2\equiv(t_2,\mathbf r_2)$, respectively.
Additionally, we use the Fourier-space variables $ k =\frac{ k_1 - k_2}{2} $ and $  q=  k_1 +  k_2$ to denote the energy-momenta conjugate to the relative coordinates $(t_{rel}, \, \mathbf r_{rel})$ and the centre-of-mass coordinates $(t, \,  \mathbf r )$, respectively. In order to simplify notations, we will use the shorthand symbols $ k $ and $q$ for $(\omega_k, \, {\mathbf k})$ and $( \Omega \equiv \omega_q, \,\mathbf q)$, respectively.

For our system, the explicit expressions for the bare retarded and advanced Green's functions at equilibrium are given by
\begin{align}
\label{listgf}
G_{bare}^{R}( \omega_k ,\mathbf{k}) &= 
{\mathcal G}^{(0)}(i\,k_0,\mathbf k) \Big \vert_  {i\,k_0 \rightarrow \omega_k +i\,0^+}
= \dfrac{1} {\omega_k- \delta_k + i \,0^+ }\,,\quad
G_{bare}^{A }(\omega_k ,\mathbf{k}) 
 = {\mathcal G}^{(0)}(i\,k_0,\mathbf k) \Big \vert_  {i\,k_0 \rightarrow \omega_k - i\,0^+} 
 =  \dfrac{1} {\omega_k- \delta_k - i \,0^+ } \,,
\end{align}
for the fermions, and
\begin{align}
\label{listgfbos}
D_{1}^{R}(\omega_q,\mathbf{q}) 
& =\mathcal D^{(1)}(i\,q_0,\mathbf q) \Big \vert_  {i\,q_0 \rightarrow \omega_q +i\,0^+}
= \frac{ -1}
{ q_{\parallel}^2
- i\,\frac{ e^2 \,m\,\omega_q } { \pi\,v_F\,| q_\parallel |}}
\,,\quad
D_{1}^{A}(\omega_q,\mathbf{q}) 
=\mathcal D^{(1)}(i\,q_0,\mathbf q) \Big \vert_  {i\,q_0 \rightarrow \omega_q - i\,0^+}
= \frac{-1}
{ q_{\parallel}^2
+ i\,\frac{ e^2 \,m\,\omega_q } { \pi\,v_F\,| q_\parallel |}}\,,
\end{align}
for the bosons. These follow straightforwardly from Eqs.~\eqref{eqbareGF} and \eqref{eqbareD}.
While performing the analytic continuation $i\,q_0 \rightarrow \omega_q +i\,0^+$,
we have used the relation
$$\text{sgn}(q_0) \equiv \text{sgn}\Big(\text{Im}(i\,q_0) \Big)
\rightarrow \text{sgn}\Big(\text{Im} \big (\omega_q+ i \,0^+ \big ) \Big)
=\text{sgn} \big ( 0^+ \big ) =1\,, $$ and
an analogous relation for the case $i\,q_0 \rightarrow \omega_q - i\,0^+$ to derive $D_{1-loop}^{A}$.
The equilibrium expressions for the retarded and advanced fermion self-energy at one-loop order, obtained from the analytic continuation of Eq.~\eqref{selfenergyform} to real frequencies, take the forms:
\begin{align}
\label{sigmara}
\Sigma^{R}(\omega_k)
= - \frac{e^{4/3} \left[ \sqrt 3\,\text{sgn}(\omega_k) + i \right]
|\omega_k |^{2/3} } 
{ 4\, v_F\,\pi ^{2/3} \left( m/v_F\right)^{1/3} } \text{  and  }
\Sigma^{A}(\omega_k)
= - \frac{e^{4/3} \left[ \sqrt 3\,\text{sgn}(\omega_k) - i \right]
|\omega_k |^{2/3} } 
{ 4\, v_F\,\pi ^{2/3} \left( m/v_F\right)^{1/3} }\,,
\end{align}
respectively.
These lead to the one-loop corrected Green's functions
\begin{align}
\left[ G_1^{R/A}(\omega_k, \mathbf{k}) \right]^{-1}
= \left[ G_{bare}^{R/A}( \omega_k ,\mathbf{k})  \right]^{-1} -\Sigma^{R/A}(\omega_k)\,.
\end{align}
The equilibrium Green's functions are related to the spectral function $A$ and the Fermi-Dirac distribution $ f_0(\omega) = \frac{1} {1 + \, e^{\beta \omega}} $ (at a temperature $T =1/\beta$) as
\begin{align}
G^<(\omega_k,\mathbf k)&=i \, f_0(\omega_k) \, A( \omega_k,\mathbf k) \,,\quad
G^>(\omega_k,\mathbf k) =-i \left [1- f_0(\omega_k) \right ] A(\omega_k,\mathbf k)\,,
 \end{align}
where
 \begin{align}
A( \omega_k,\mathbf k) = -  2\, \text{Im}[G^R(\omega_k,\mathbf{k})]
=\dfrac{ 2\, \text{Im} [\Sigma^R( \omega_k,\mathbf k)]}
{ \Big [ \,\omega_k- \delta_k-\text{Re} [\Sigma^R ( \omega_k,\mathbf k)] \,\Big]^2
+ \Big [ \, \text{Im} [\Sigma^R ( \omega_k,\mathbf k)] \,\Big]^2} \,.
 \end{align}

In Landau's Fermi liquid theory, the imaginary part of the fermionic self-energy turns out to be $\text{Im}[\Sigma^R] \sim \omega_k^2 \ll |\omega_k| $ for $|\omega_k| \rightarrow 0 $.
As a result, the equilibrium spectral function $A$ takes the form of a sharply-peaked function of $\omega_k$, such that $ A(\omega_k,\mathbf{k}) 
\simeq 2 \,\pi\, \delta \big( \omega_k - \xi_\mathbf{k} - \text{Re}[\Sigma^R(\omega_k,\mathbf{k})] \big)\,,$
where $ \xi_\mathbf{k}$ is the bare fermion dispersion.
This relation indicates that for fluctuations close to the equilibrium, we can construct a closed set of  equations for the fermion distribution function $ f(\omega_k,\mathbf k;t,\mathbf r)$,
which constitute the QBEs. The set of linearized QBEs for the fluctuation
$ \delta f(\omega_k,  \mathbf k;t,\mathbf r) = f(\omega_k, \mathbf k;t,\mathbf r) - f_0(\omega_k)$ describes the transport equations for a Fermi liquid.

For our NFL system, Eq.~\eqref{sigmara} gives $ \text{Im} [\Sigma^R] \propto |\omega_k |^{2/3}$, implying that $A(\omega_k, \mathbf{k})$ is not a sharply-peaked function of $\omega_k $ at equilibrium, in contrary to the behaviour of Fermi liquid systems. As a result, $  \delta f(\omega_k, \mathbf k;t,\mathbf r)$ does not satisfy a closed set of equations even at equilibrium. We thus need to devise a formalism which does not depend on the smallness
of the decay rate, a quantity that is proportional to the width of the peak in $A (\omega_k, \mathbf{k}) $ as a function of $\omega_k$. Observing that $A (\omega_k, \mathbf{k})$ has a well-defined peak around $ \delta_k = 0$ \cite{prange,kim_qbe} (since $\Sigma^R$ is a function of $\omega_k $ only), and $\int_{-\infty}^{\infty} \frac{d\delta_k}{ 2\, \pi  } A (\omega_k, \mathbf{k}) = 1$, 
we conclude that $G^<$ and $G^>$ are sharply-peaked functions of $\delta_k$. Integrating over the region of the sharp peak, it is useful to define the generalized fermion distribution function $f$ (also known as a Wigner distribution function) as
\begin{align}
\label{eq:gdf}
\int \frac{d\delta_k}{ 2\, \pi  } \,
G^<(\omega_k, {\mathbf k} ; \omega_q,\mathbf q) 
= i\,f(\omega_k, {\mathbf k} ;\omega_q, \mathbf q)\,,\quad
 \int \frac{d\delta_k }{ 2\, \pi  } \,
G^>(\omega_k, {\mathbf k} ; \omega_q,\mathbf q) 
= 
i   \left[ f(\omega_k,{\mathbf k};\omega_q, \mathbf q) -1\right] ,
\end{align}
which works in the absence of well-defined Landau quasiparticles. These relations will allow us to derive the set of QBEs which can characterize the fluctuations of a critical Fermi surface, as long as the system is not far away from the equilibrium.

\section{QBE for fermions}

\label{secqbe}

In order to derive the QBEs for the collective modes of a critical Fermi surface, we need to refer to its global properties. For the sake of simplicity, we assume a circular Fermi surface with the Fermi momentum vector give by  ${\mathbf k}_F = k_F \, \hat{{\mathbf k}}_{\text{rad}}$, where $ \hat{{\mathbf k}}_{\text{rad}}$ is the angle-dependent unit vector pointing radially outward on the Fermi surface. Moreover, our derivations will be limited to the zero temperature limit (i.e., $T=0$).
We use the parametrization $\mathbf {k} =  \frac{\mathbf k_1 -\mathbf k_2} {2} 
\equiv \left( k_F + k_\perp \right) \hat{{\mathbf k}}_{\text{rad}} 
+ k_\parallel \,{\hat{ \mathbf {\theta}}}_{\mathbf k} $, $\mathbf q = \mathbf k_1 + \mathbf k_2$,  and $\theta_{\mathbf k} \,
( \theta_{\mathbf q} )$ is the angle that the vector $\mathbf {k} \,(\mathbf q)$ makes with the $x$-axis. Since we are focussing on small perturbations of the Fermi surface located at momentum $k_F$, we must have $ \left \lbrace |k_\perp|, |k_\parallel|,
|\mathbf q |  \right \rbrace \ll k_F$. The functional dependence of $f(\omega_k, \mathbf k ;\omega_q, \mathbf q)$ on $\mathbf k$ thus effectively reduces to the angle $\theta_{\mathbf q  \mathbf k} = \theta_{\mathbf q}-\theta_{\mathbf k} $ (i.e., the angle between ${\bf k}$ and ${\bf q}$), which we symbolically express by using the notation $f(\omega_k, \theta_{\mathbf q  \mathbf k} ;\omega_q, \mathbf q)$.

Not too far away from from the equilibrium, the equations of motion for the fermionic operators lead to the linearized QBEs for the fermions \cite{prange,kim_qbe,ips-zero-mode} as follows:
 \begin{align}
   &  \left[G_0^{-1}( k +  q/2)-G_0^{-1}( k- q/2)\right] \delta G^<(k,q)
-\Big [ \text{Re} [ \Sigma_0^R( k + q/2) ]  -\text{Re}  [\Sigma_0^R(k-q/2)]
\Big ] \, \delta G^<(k,q)\nonumber \\ 
+  &  \, \left[G_0^<( k + q/2)-G_0^<( k- q/2)\right]
\delta \Big (\text{Re} [ \Sigma^R(k,q)]\Big )
-\left[\Sigma_0^<( k+q/2)
-\Sigma_0^<( k-q/2)\right]\delta \Big (\text{Re} [ G^R(k,q)] \Big ) \nonumber \\
+    &  \, \Big [\text{Re} [ G_0^R( k+q/2)]-\text{Re} [ G_0^R( k-q/2)]
\Big] \, \delta \Sigma^<(k,q) = I_{coll} (k,q)\,,
\label{QBEf}
 \end{align} 
 where
$\delta { G} = {  G} - { G}_0$ and
$\delta {\Sigma} = { \Sigma} - { \Sigma}_0$ 
denote the small deviations from equilibrium, and
 \begin{align}
 I_{\text{coll}} (k, q) = 
    G_0^<(k)\, \delta \Sigma^>(k,q) + \Sigma_0^>(k)  \,\delta G^<(k,q)
    -G_0^>(k)  \,  \delta \Sigma^<(k,q)-\Sigma_0^<(k)   \,   \delta G^>(k,q)  \,.
\label{collf}
 \end{align}
gives the collision integral.
The equilibrium Green's functions and self-energies have been indicated by using the subscript ``$0$''. 
We do not repeat here the intermediate steps for the derivation, which can be found in Refs.~\cite{prange,kim_qbe,ips-zero-mode}. 
As in Ref.~\cite{ips-zero-mode}, the QBEs for the fermions have been obtained by assuming that the bosons are always in equilibrium, such that we always use the boson distribution function 
$ n_b(\nu)  = {1} /( e^{\beta \, \nu}-1 ) $, and the bosonic Green's functions do not depend on the relative coordinates $(t_{rel}, \mathbf r_{rel})$. We also point out that while using the forms of the boson propagators from Eq.~\eqref{listgfbos}, $D^R(\nu, \mathbf p-\mathbf p')$ depends on the component of the exchange momenta $\mathbf p-\mathbf p'
\simeq  2 \, k_F \sin (\theta_{\mathbf p \mathbf p'}/2)
\,{\hat{ \mathbf {\theta}}}_{\mathbf p} $
(between two nearby patches with almost parallel tangent vectors) projected along the Fermi surface. Hence Eq.~\eqref{listgfbos} can approximated by
 \begin{align}
D_{1}^{R}(\nu,\mathbf{p}-\mathbf{p}') 
& \simeq D_{1}^{R}(\nu, q_{\bp \bp'}) 
 =-\dfrac{1}{ q_{\bp \bp'}^2-i\, \chi \, {|\nu|} / q_{\bp \bp'}}\,,
\quad
q_{\bp \bp'} \equiv  2 \, k_F \sin ( {|\theta_{\bp'\bp}|}/ 2 ) \,, 
\nn
\chi &= \dfrac{e^2\, m} {\pi\,  v_F }
=  8 \,m \, k_F \,\alpha\,,\quad 
\alpha=\dfrac{e^2} {16 \,\pi\, E_F} 
= \dfrac{e^2} { 8 \,\pi\, v_F\,k_F}
\,,
\end{align}  
where $ \alpha $ is the fine structure constant (with $E_F = v_F\,k_F / 2$ being the Fermi energy). This implies that
\begin{align}
\text{Re} [D_{1}^R(\nu, q_{\bp \bp'})] =
 -\dfrac{q_{\bp \bp'}^4} {q_{\bp \bp'}^6+\chi^2 \,\nu^2} \,,\quad 
\text{Im} [D_{1}^R(\nu,  q_{\bp \bp'})]
=-\dfrac{ \chi  \, q_{\bp \bp'}\, |\nu|} 
{q_{\bp \bp'}^6+\chi^2 \,\nu^2}\,.
\label{reimD}
\end{align}

Some straightforward but tedious algebra gives us the following expressions for various fermion self-energies:
 \begin{align}
& \Sigma^<( \omega_{k'},\mathbf k';\Omega, \bq)/ \left[  -2\, i\, N(0)\, e^2 \right] \nn & =
  \int \dfrac{d\theta_{\bp \bq}}{2\,\pi}\int_0^\infty \dfrac{d\nu}{\pi}   \,
 \text{Im} [ D_1^R( \nu, q_{ \mathbf k' \bp})  ]
 \left[ n_b(\nu) \,f(\omega_{k'} -\nu,\theta_{\bp \bq};\Omega,\bq) 
 +\left \lbrace n_b(\nu)+1   \right \rbrace 
  f(\omega_{k'} +\nu, \theta_{\bp \bq}; \Omega, \bq) \right] ,
\nn & \Sigma^>(\omega_{k'},\mathbf k';\Omega, \bq) / \left[  -2\, i\, N(0)\, e^2 \right]
\nn & =
 \int \dfrac{ d\theta_{\bp \bq}} {2\,\pi}
 \int_0^\infty \dfrac{d\nu}{\pi} 
\, \text{Im} [ D_1^R(\nu, q_{ \mathbf k' \bp}) ]  
 \left [  n_b(\nu) \left \lbrace f( \omega_{k'} + \nu, \theta_{\bp \bq}; \Omega, \bq)-1 \right \rbrace  
 +\left \lbrace n_b(\nu)+1 \right \rbrace  
 \left \lbrace f( \omega_{k'}-\nu, \theta_{\bp \bq}; \Omega, \bq)
 -1 \right \rbrace  \right ] ,
\nn 
& \text{Re}  [\Sigma^R(\omega_{k'}, \mathbf k';\Omega,\bq) ]  =- e^2\, N(0)
 \int \dfrac{d\theta_{\bp \bq}} {2 \,\pi}
 \int_0^\infty \dfrac{d\nu}{\pi} \, 
\text{Re} [ D_1^R(\nu-\omega_{k'},q_{ \mathbf k' \bp }) ]
 \,f(\nu,\theta_{\bp \bq};\Omega,\bq)  \,.
 \end{align}
which need to be plugged into Eq.~\eqref{QBEf}.
Here, $N(0)$ is the fermionic density of states at the Fermi surface.

Using the above expressions along with Eq.~\eqref{eq:gdf} and integrating over $\delta_k / (2\,\pi)$, we get 
 \begin{align}
   & i  \left(  \Omega- v_F   \,  |\mathbf q| \cos \theta_{\mathbf k \bq} \right)
   \delta f(\omega_k,\theta_{\mathbf k  \bq};\Omega,\bq) \nonumber \\
 & +i \,e^2 \,N(0)
  \int \dfrac{d\theta_{\bp \bq}} {2 \,\pi}
 \int_{-\infty}^\infty \dfrac{d\nu}{\pi} \, 
 \text{Re} [ D_1^R(  \nu-\omega_k, q_{ \mathbf k \bp  } ) ]
 \left[  \Theta( -\nu - {\Omega}/{2})- \Theta( - \nu + {\Omega}/{2})  \right ]
 \delta f(\omega_k ,\theta_{\mathbf k  \bq};\Omega,\bq) \nonumber \\
   & -i \,e^2 \,N(0)
  \int \dfrac{d\theta_{\bp \bq}} {2 \,\pi}
 \int_{-\infty}^\infty \dfrac{d\nu}{\pi} \, 
\text{Re}  [ D_1^R(   \nu-\omega_k, q_{ \mathbf k \bp}) ]
\left[  \Theta(-\omega_k- {\Omega}/{2})
 -\Theta( -\omega_k + {\Omega}/{2}) \right ]
\delta f(\nu,\theta_{\bp \bq};\Omega,\bq) 
\end{align}
for the left-hand-side of Eq.~\eqref{QBEf}, and
\begin{align}
 \int \dfrac{d   \delta_k}{ 2\, \pi } \, I_{coll} (k,q) 
/ [-2\, e^2 \,N(0)]  &
\nn  =   \int_{-\infty}^{\infty} d\omega_p 
\int \dfrac{d\theta_{\bp \bq}}{ 2\, \pi }
  \int_0^\infty \dfrac{d\nu}{\pi}  \,
& \Big[ \Big \lbrace \delta f(\omega_k, \theta_{ \mathbf k \bq}; \Omega, \mathbf q)
 \Big( 1+n_b(\nu)- \Theta(-\omega_p) \Big )
 -\delta f(\omega_p,\theta_{\bp \bq} ;\Omega, \mathbf q)
 \Big  ( n_b (\nu)+ \Theta(-\omega_p) \Big )\Big \rbrace 
\nn &
\hspace{ 0.5 cm }
 - \Big \lbrace
 \delta f(\omega_p,\theta_{\bp \bq} ; \Omega, \mathbf q)
 \left( 1+n_b (\nu)- \Theta(-\omega_k)\right )
 -\delta f(\omega_k, \theta_{\bp\bq}; \Omega, \mathbf q)
 \Big ( n_b (\nu)+ \Theta(-\omega_p) \Big) \Big  \rbrace  \Big ]
\nn 
& \hspace{ 0.2 cm }
\times  \text{Im}  [D_1^R(\nu, q_{\mathbf k \bp})]
\,   \delta(\omega_p-\omega_k +\nu)\,. 
 \end{align}
Finally, we integrate over $ \omega_k$ to obtain
 \begin{align}
 \label{final_eq}
 \left (  \Omega - v_F \,|\mathbf q| 
 \cos{\theta_{\mathbf k  \bq}
 }\right ) u(\theta_{\mathbf k \bq};\Omega,\bq)
 +I_1 = I_2 \,,
\text{ where }
u(\theta_{\mathbf k \bq};\Omega,\bq)=
\int_{-\infty}^{\infty} \dfrac{d\omega_k }{ 2\, \pi  } 
\,  \delta f( \omega_k , \theta_{ \mathbf k \bq}; \Omega, \bq)\,,
\end{align}
\begin{align}
 & I_1/[e^2 \, N(0)] \nn
 & = 
  \int_{-\infty}^{\infty} \dfrac{d  \omega_p}{\pi}
 \int_{-\infty}^{\infty} \dfrac{d\omega_k }{ 2\, \pi  } 
 \int \dfrac{d\theta_{\bp \bq}} { 2\, \pi } 
\left[\delta f(\omega_k,\theta_{ \mathbf k \bq};\Omega,\bq)
 -\delta f(\omega_k ,\theta_{\bp   \bq};\Omega,\bq)\right]
 \text{Re} [D_1^R( \omega_p -\omega_k,  q_{\bp  \mathbf k})]
 \left[ \Theta ( -\omega_p - {\Omega}/2)-\Theta( -\omega_p  + \Omega /2)\right] 
\nn
& = -\text{sgn}(\Omega)
 \int \dfrac{d\theta_{\bp'\bq}} {2\,\pi}  
 \int_{-\infty}^{\infty} \dfrac{d\omega}  {2 \, \pi}  
  \left[\delta f(\omega,\theta_{\bp\bq};\Omega,\bq)-\delta f(\omega,\theta_{\bp'\bq};\Omega,\bq)\right]
  \int_{-|\Omega|/2}^{|\Omega|/2} \dfrac{d  \omega_p}{\pi}
\,  \text{Re} [D_1^R(  \omega_p -\omega, q_{\bp'\bp}) ]\, ,
\label{I1}
\end{align}
and
\begin{align}
& I_2 /[ 2 \, i \,e^2 \,N(0) ]
\nn  & = \int_{0}^{\infty} \dfrac{d\nu}  {\pi}
  \int_{-\infty}^{\infty} d  \omega_p \int_{-\infty}^{\infty} \dfrac{d\omega_k} {2\,\pi}
 \int \dfrac{d\theta_{\bp   \bq}}   { 2\, \pi } 
  \, \text{Im} [ D_1^R ( \nu,q_{\bp \mathbf k }) ]
\left[
  \delta f(\omega_k ,\theta_{ \mathbf k    \bq};\Omega,\bq)
  -\delta f(\omega_k ,\theta_{\bp \bq};\Omega,\bq)
\right]  \nonumber \\
  &  \hspace{ 4.5 cm} \times  
\left[  \delta(\omega_p-\omega_k +\nu) 
  \left \lbrace   1-\Theta(-\omega_p) +   n_b(\nu)\right \rbrace 
 +  \delta(\omega_p -\omega_k-\nu)
  \left \lbrace \Theta(-\omega_p)+n_b(\nu)   \right \rbrace  
\right ] \nn
& =  \int_{0}^{\infty} \dfrac{d\nu}{\pi} 
\int_{-\infty}^{\infty} \dfrac{d\omega_k}{2 \, \pi} 
\int \dfrac{d\theta_{\bp'\bq}} { 2\, \pi }  
\, \text{Im} [D_1^R( \nu, q_{\bp'\bp\bq})]
   \left[\delta f(\omega_k,\theta_{\bp\bq};\Omega,\bq)
   -\delta f(\omega_k ,\theta_{\bp'\bq};\Omega,\bq)\right] 
\nonumber \\  &  \hspace{ 4.5 cm} \times     
\left [
\int_{-\infty}^0 d\omega_p \,\delta(\omega_p-\omega_k -\nu) 
 + \int_0^{\infty} d\omega_p \,\delta(\omega_p-\omega_k +\nu) \right ]
\nn & =
\int_{-\infty}^{\infty} \dfrac{d\omega_k}{2 \, \pi} 
\int \dfrac{d\theta_{\bp'\bq}}{ 2\, \pi }  
  \left[\delta f(\omega_k,\theta_{\bp\bq};\Omega,\bq)-\delta f(\omega_k,\theta_{\bp'\bq};\Omega,\bq)\right]  
  \left[ \Theta(\omega_k)  
  \int_{0}^{\omega_k} \dfrac{d\nu}{\pi} \,
 \text{Im} [D_1^R( \nu, q_{\bp'\bp} )]
 +\Theta(-\omega_k)
  \int_{0}^{|\omega_k|} \dfrac{d\nu}{\pi} \,
  \text{Im} [D_1^R( \nu, q_{\bp' \bp}) ] \right] .
\label{I2} 
 \end{align} 
Here, we have used the fact that the boson distribution function $n_b(\nu)=0$ at $T = 0$.
We will focus on the $\Omega \geq 0$ regions, without any loss of generality, to simplify the expressions.
 
\subsection{Simplifying the integrals}
\label{secsimplify}

Eq.~\eqref{I1} can be rewritten as
\begin{align}
 I_1& = \text{sgn}(\Omega)\,e^2 \, N(0) \, \Omega^2
\int \dfrac{d\theta_{\bp'\bq}} {2\,\pi}  \int_{-\infty}^{\infty} 
\dfrac{d \tilde y}{2 \, \pi} 
  \left[\delta f( \tilde y \, |\Omega|,\theta_{\bp\bq};\Omega,\bq)
 -\delta f(\tilde y\,|\Omega|,\theta_{\bp'\bq};\Omega,\bq)\right]\,
 \tilde I_1\,,\nn
 \tilde I_1 & =  \int_{-1/2}^{1/2} \dfrac{d \tilde x}{\pi} 
  \dfrac{ q_{\bp'\bp}^4}
{ q_{\bp'\bp}^6 + \chi^2 \,\Omega^2\,(\tilde x-\tilde y)^2}\,.
\end{align}
Observing that $\tilde I_1$ is peaked at $\tilde y=0$, we can simplify it as 
 \begin{align}
\tilde I_1
\simeq  \int_{-1/2}^{1/2} \dfrac{d \tilde x} {\pi} 
\dfrac{  q_{\bp'\bp}^4}
{  q_{\bp'\bp}^6 + \chi^2 \, \Omega^2 \,\tilde x^2}
=  \dfrac{ 2\, q_{\bp'\bp}} {\pi \, \chi \, \Omega} \,
\tan^{-1}\left({\dfrac{\chi\, \Omega}{2 \,  q_{\bp'\bp}^3}}\right).
 \end{align}
The advantage of setting $\tilde y=0$ in the above integration is that it allows us to get an an approximate analytical expression. Furthermore, we can now use the relation
$
 |\Omega| \int_{-\infty}^{\infty} \dfrac{d \tilde y} {2 \, \pi} 
\, \delta f(\tilde y \,|\Omega|, \theta; \Omega, \bq)= u(\theta;\Omega,\bq)
$ 
to get the simplified expression
\begin{align}
 I_1 & \simeq \dfrac{N(0) \, e^2} {\chi} 
 \int \dfrac{d\theta_{\bp' \bq}}{2 \, \pi} 
 \left[
 u(\theta_{\bp \bq};\Omega, \bq)-u(\theta_{\bp' \bq};\Omega, \bq)
 \right] 
 \dfrac{2 \,q_{\bp'\bp}} {\pi}
\,\tan^{-1}\left({\dfrac{\chi \, \Omega}{2 \, q_{\bp'\bp}^3}}\right) 
\nn &
=k_F \,v_F \int \dfrac{d\theta_{\bp' \bq}}{2 \, \pi} \left[u(\theta_{\bp \bq};\Omega, \bq)-u(\theta_{\bp' \bq};\Omega, \bq)\right] F^R(\theta_{\bp' \bp},a)
 \,,\nn
F^R(\theta,a)
& =\dfrac{2}{\pi} \tan^{-1}\bigg(\dfrac{a}
{2 \sin^3 ({\theta} /{2})}  \bigg) \sin ({\theta} / {2} ) \,,\quad
a=\dfrac{\chi \,\Omega}{8 \,k_F^3} \,,
 \label{I1final}
 \end{align}
where we have used $N(0) = \dfrac{m}{2 \, \pi}$.
We will refer to the function $F^R(\theta,a)$ as the real part of the generalized Landau parameter.

 In order to simplify Eq.~\eqref{I2}, we observe that 
$ \text{Im} [D_1^R( \nu, q_{\bp' \bp}) ]$ is an even function of the frequency $\nu$, which leads to
\begin{align}
 I_2 &= 2 \, i \,e^2 \,N(0) 
\int_{-\infty}^{\infty} \dfrac{d\omega_k}{2 \, \pi}  
 \int \dfrac{d\theta_{\bp'\bq}}{ 2\, \pi }  
  \left[\delta f(\omega_k,\theta_{\bp\bq};\Omega,\bq)
 -\delta f(\omega_k,\theta_{\bp'\bq};\Omega,\bq)\right] \,\tilde I_2\,,\nn
\tilde I_2
& = \int_{0}^{|\omega_k|} \dfrac{d\nu}{\pi}  
\, \text{Im}[D_1^R( \nu, q_{\bp'\bp})]
= -\dfrac{  q_{\bp'\bp}} {2\, \pi \,\chi} 
\, \ln \Big(1+{\chi^2 \,\omega_k^2} /   q_{\bp'\bp}^6 \Big ) \,.
\end{align} 
Since log is a slowly varying function, we can approximate the argument of log in $\tilde I_2$ by setting $\omega_k \simeq \Omega$. With this simplification, we finally obtain the expression
\begin{align}
I_2 & \simeq -\dfrac{2 \,i \,e^2 \, N(0) \,k_F} {\pi \,\chi} 
\int \dfrac{d\theta_{\bp'\bq}}{ 2\, \pi }  
\left[u(\theta_{\bp\bq})-u(\theta_{\bp'\bq})\right]  \,
\sin ( |\theta_{\bp' \bp}| /{2} ) \,
\ln \Big (1+ a^2/ \sin^6 (\theta_{\bp' \bp} /2 )\Big) \nn  
& =-i\, k_F\, v_F \int \dfrac{d\theta_{\bp'\bq}}{ 2\, \pi }  
\left[u(\theta_{\bp\bq})-u(\theta_{\bp'\bq})\right] 
F^I(\theta_{\bp'\bq}-\theta_{\bp\bq},a)\,,\nn
F^I(\theta,a) & =
\dfrac{1}{\pi} \sin (|\theta|/2)
 \ln \Big(1+ a^2 / \sin^6 (\theta/2) \Big)\,.
\label{I2final}
\end{align}
We will refer to the function $F^I(\theta,a)$ as the imaginary part of the generalized Landau parameter.

 \subsection{Final form of QBE}
\label{secfinalqbe}
 
Using Eqs.~\eqref{I1final} and \eqref{I2final}, the QBE shown in Eq.~\eqref{final_eq}
reduces to
  \begin{align}
 \left(\bar{\Omega}- \bar{q} \cos\theta\right) u(\theta)
+ \int \dfrac{d \tilde \theta} { 2\, \pi }  
\left[u(\theta)-u( \tilde \theta)\right] F(\theta-\tilde \theta,a) =0\,, \quad
F (\theta,a) \equiv F^{R} (\theta,a) + i\, F^I (\theta, a)\,,
 \label{QBE1}
  \end{align}
 where $\bar{\Omega}= \Omega / (v_F \,k_F)$, $\bar{q}= |\mathbf q|/ k_F$, $\theta\equiv\theta_{\bp \bq}$, $ \tilde \theta \equiv\theta_{\bp' \bq}$, and $u(\theta)\equiv u(\theta;\Omega,\bq)$. The dimensionless parameter $a$ can be written as $ a=\bar{\Omega}\, \alpha$. We decompose $u(\theta)$ and $F (\theta) \equiv F^{R} (\theta) + i\, F^I (\theta)$ into angular momentum components as follows:
 \begin{align}
 \label{eq_decomp}
& u(\theta) = u^+(\theta) + u^-(\theta)\,,\quad
u^+(\theta) =  u_0^+
  +2 \, \sum_{\ell=1} ^\infty  u_{\ell} ^
  + \cos(\ell \, \theta )\,,\quad
 u^-(\theta)  = 2  \,\sum_{\ell=1} ^\infty  u_{\ell} ^- \sin(\ell \, \theta )\,,
 \nn
& F (\theta,a) = F_0(a) + 2  \,\sum_{\ell=1} ^\infty  F_\ell (a) \cos(\ell \, \theta )\,,
\quad F_{\ell}(a) =  F_{\ell}^R(a)+i \, F_{\ell}^I(a)\,,
 \end{align}
where $ u_{\ell} ^+$ and $  u_{\ell} ^-$ represent the coefficients of the decomposition representing the parts which are even and odd, respectively, under $\theta \rightarrow  -\theta $. We note that the coefficients for the decomposition of $F(\theta,a)$ are even and therefore, $ F_{\ell}^R$ and $F_{\ell}^I $ are real functions.

Using the angular momentum decomposition coefficients, the QBE reduces to two sets of infinite number of coupled equations:
 \begin{align}
\label{zero_mode_even}  
& \bar{\Omega}\,u_0^+-\bar{q} \,u_1^+ = 0 \,,\quad
 u_{\ell} ^+-  \dfrac{ 2\,\bar{\Omega}}{\bar{q}}
  \left(1+\dfrac{F_0-F_{\ell-1}} {\bar{\Omega}}\right) u_{\ell-1}^+
 + u_{\ell-2}^+ = 0 \text{ for } \ell \geq 2  \,,
\end{align} 
and
\begin{align}
\label{zero_mode_odd}
  \bar{\Omega}  \,u_1^- -\dfrac{\bar{q}\, u_2^-} {2} 
  +\left(F_0- F_1^R \right)u_1^- = 0 \,,\quad
  \bar{\Omega} \, u_{\ell} ^-
 -\bar{q}\, \frac{ u_{\ell+1}^-+ u_{\ell-1}^- }   {2}
  + u_{\ell} ^-\left(F_0- F_{\ell} \right)= 0 
 \text{ for } \ell \geq 2\,.
  \end{align}
These recursive equations are of the form of a one-dimensional non-Hermitian tight-binding model, with the the angular momentum channels playing the role of the lattice sites.

\section{Fermi surface displacements in the collisionless limit}

 \label{secsoln}
 
  \begin{figure}[]
 \centering
   \subfigure[]{\includegraphics[width= 0.45 \textwidth]{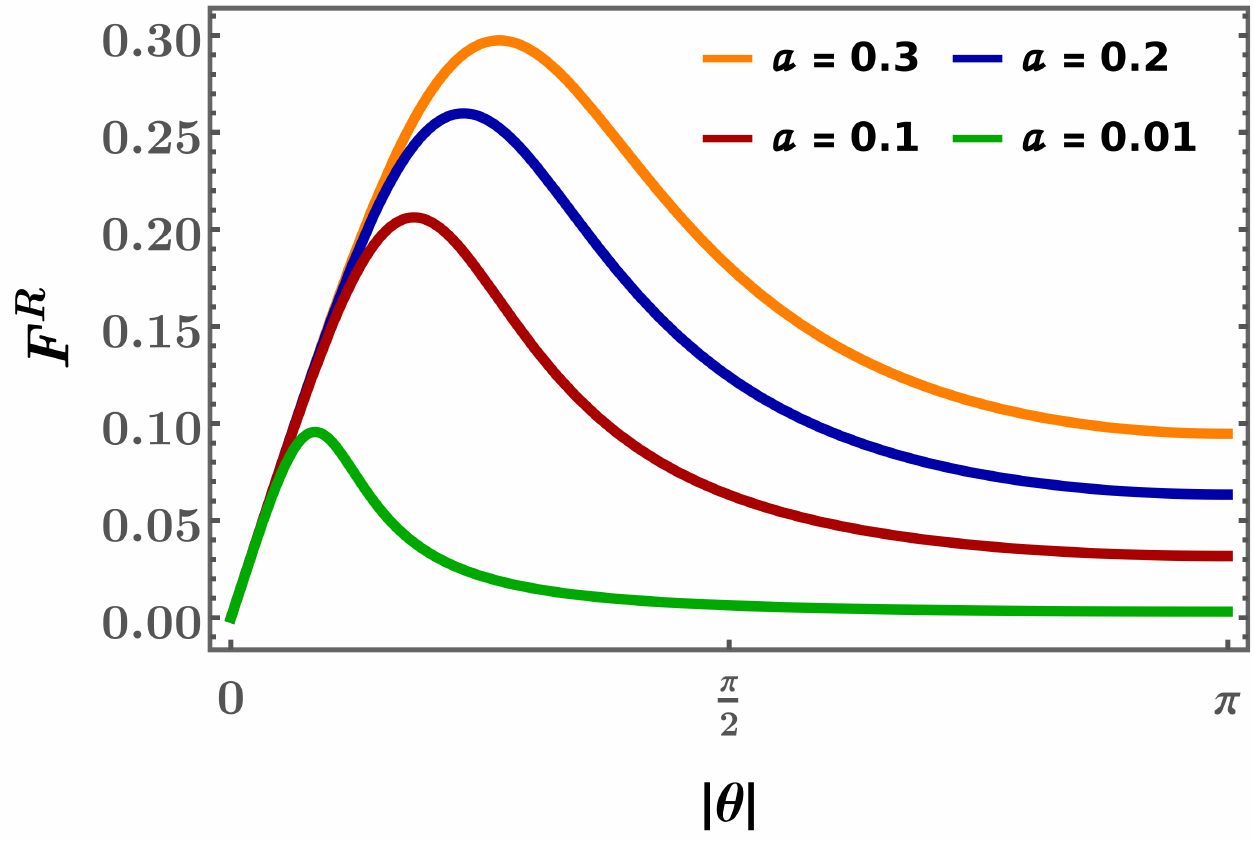}}\hspace{1 cm}
   \subfigure[]{\includegraphics[width= 0.45 \textwidth]{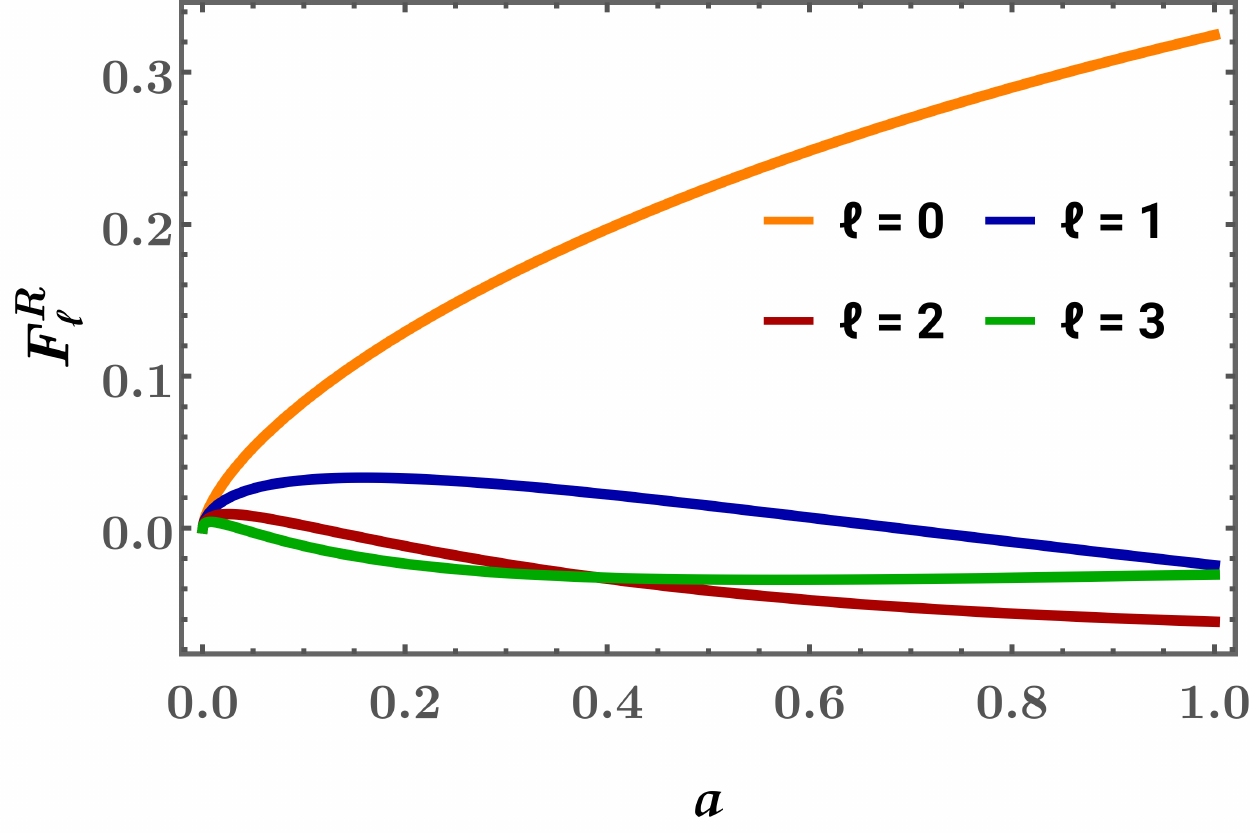}}
\caption{\label{Landau}
(a) Behaviour of the real part of the generalized Landau parameter as a function of $ \theta$, which parametrizes the angular position on a circular Fermi surface.
(b) Behaviour of the coefficients $F^R_\ell$ obtained from the decomposition of $F^R$
in angular momentum channels labelled by $\ell$, as functions of the paramater $a$.
}
 \end{figure}

In this paper, we will investigate the behaviour of the generic Fermi surface displacements in the collisionless limit, i.e., by setting $F^I = 0 $. This is a reasonable thing to do as we have assumed that the critical bosons are always at equilibrium and, therefore, the fermions have nowhere to dump their momentum. Hence, keeping a nonzero $F^I$ is inconsistent with this picture. In other words, assuming that the bosons are at equilibrium is an approximation which makes the calculations a bit easier, and this assumption is consistent with the second approximation of setting $F^I = 0 $ (as a nonzero $F^I $ would be responsible to cause the decay of the fermionic collective modes).

In Fig.~\ref{Landau}, we have plotted the behaviour of $F^R(\theta,a)$ and $F^R_\ell(a)$.
We find that $F^R_0 \gg  F^R_{\ell>0} $. 
Using this observation, we first try to understand physically the nature of the excitation modes on the Fermi surface by assuming that only $F^R_0 \neq 0$ (i.e., we set $ F^R_{\ell>0} =0$). Our aim is to understand how the boundary between the well-defined collective modes and the particle-hole continuum is influenced by the complicated frequency dependence of the $F^R$ parameter. We also want to figure out the dispersion of the collective mode(s). For the sake of comparison, we would like to remind the reader that for a Fermi liquid, the dispersion of the zero sound collective mode is $\Omega = \left( v_F \, \sqrt{ {\mathcal F}_0 / 3 } \right) |\mathbf q|$ (where ${\mathcal F}_\ell $ is the coefficient of the $\ell^{\text{th}}$ angular momentum channel decomposition of the Landau parameter), in a simple model where only ${\mathcal F}_0 $ is taken to be nonzero \cite{pines}.

\subsection{Even parity sector}
\label{seceven}

  \begin{figure}[]
  \centering
 \subfigure[]{\includegraphics[width = 0.45 \textwidth]{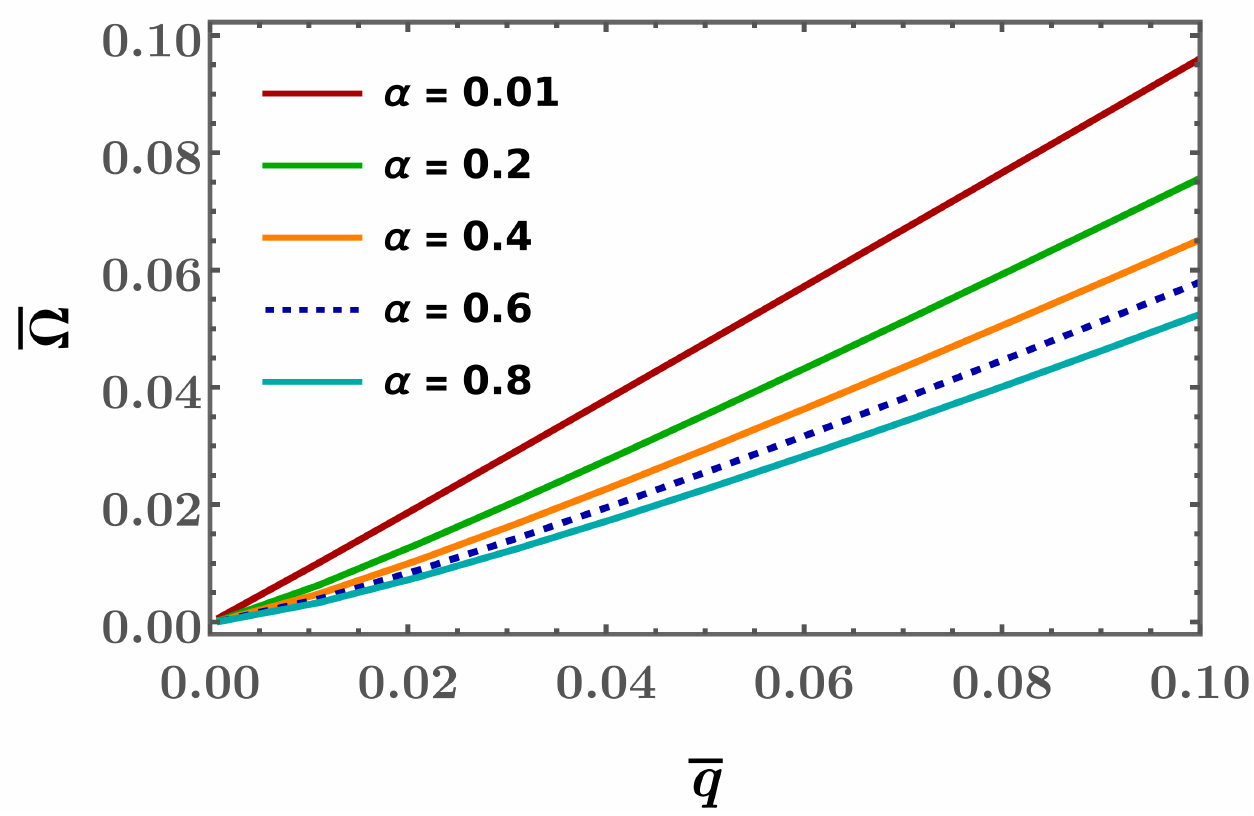}} \hspace{1 cm}
 \subfigure[]{\includegraphics[width = 0.47 \textwidth]{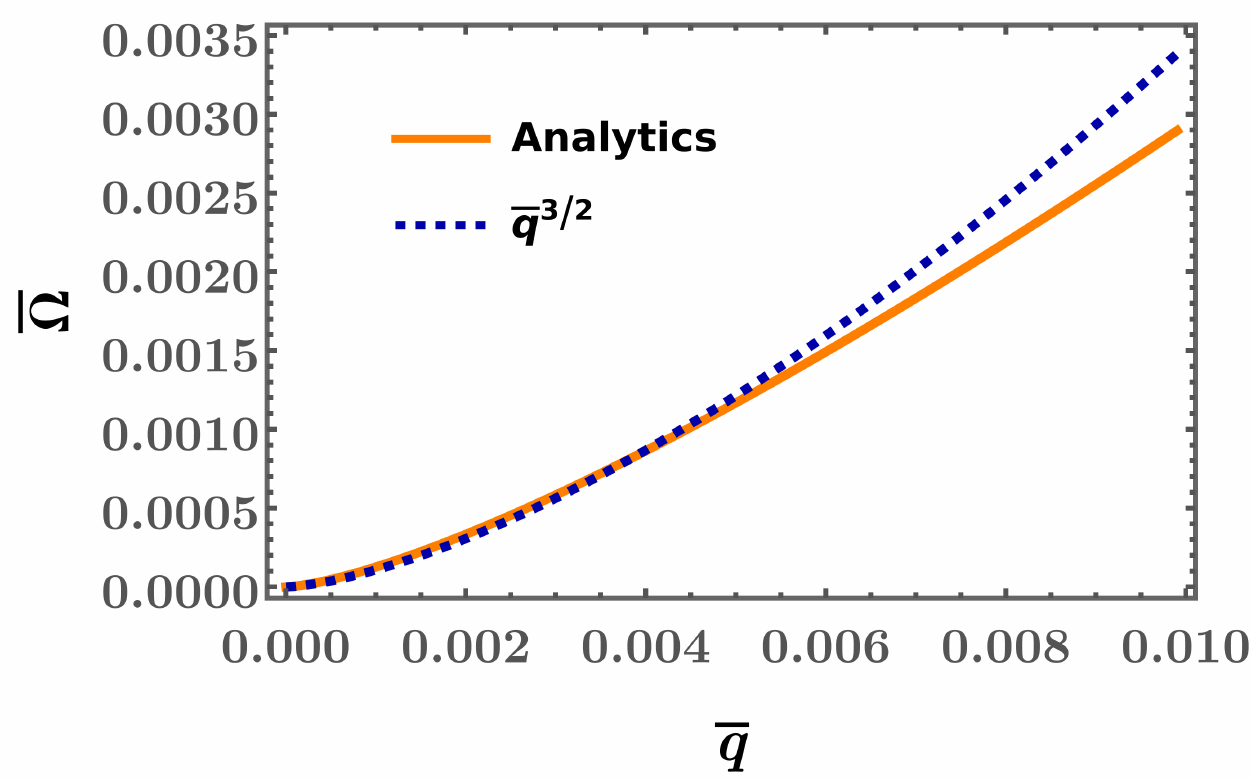}} 
 \subfigure[]{\includegraphics[width = 0.45 \textwidth]{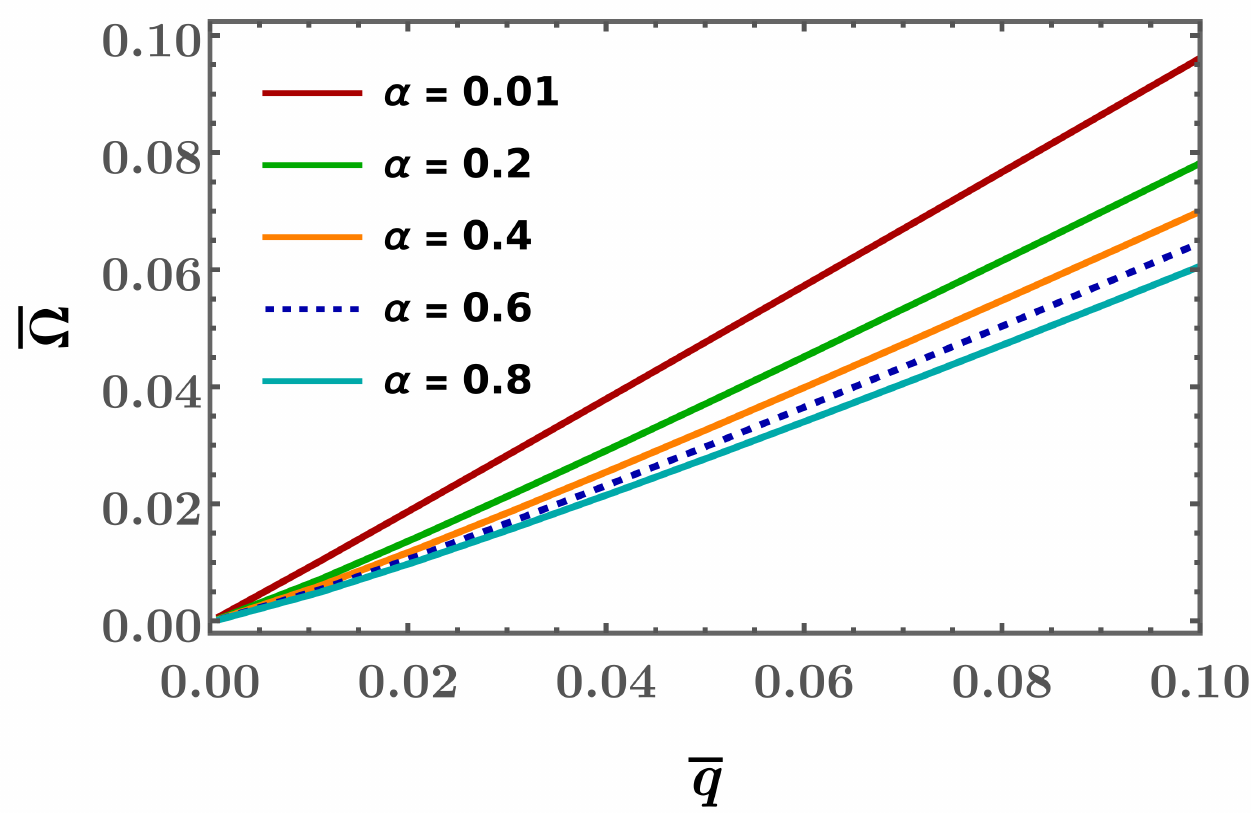}} \hspace{1 cm}
 \subfigure[]{\includegraphics[width = 0.47 \textwidth]{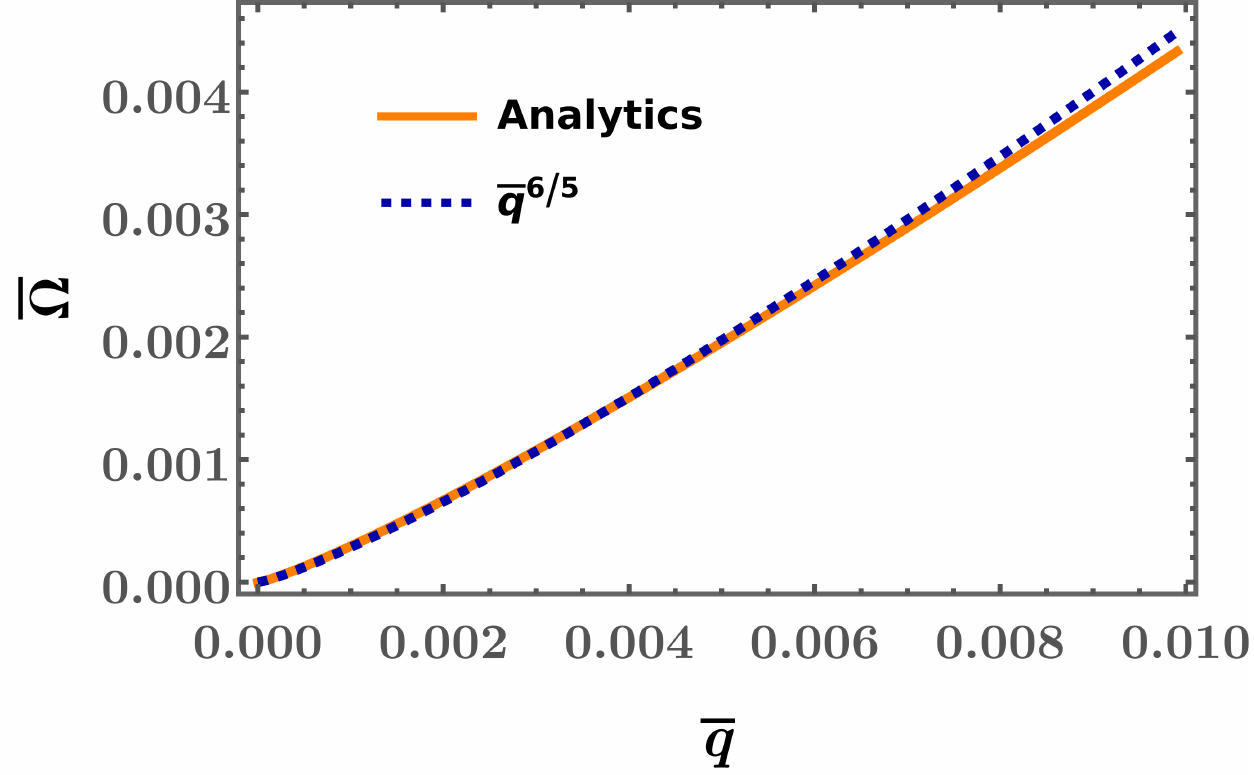}} 
  \caption{Boundary for the dispersion of the continuum of localized excitations, given by the condition $ \bar{\Omega}+ F^R_0 = \bar q $, plotted for (a) different values of $\alpha$; (b) very small values of $\bar \Omega$ with $\alpha = 0.8 $, where it shows a good fit with the behaviour $\bar \Omega \sim {\bar q}^{3/2} $.
Dispersion of the delocalized collective mode excitation of the Fermi surface, plotted for (c) different values of $\alpha$; (d) very small values of $\bar \Omega$ with $\alpha = 0.8 $, where it shows a good fit with the behaviour $\bar \Omega \sim {\bar q}^{6/5} $.
In subfigures (b) and (d), the solid orange curves represent the results obtained from our analytical expressions in Sec.~\ref{seceven}, whereas the dashed blue curves correspond to the fitting functions, each with the appropriate fractional power dependence of $\bar q$.
\label{ph_zero}
}
 \end{figure}

We first focus on the even parity solutions of Eq.~\eqref{QBE1}, given by $u_\ell^+$ of Eq.~\eqref{eq_decomp}. We need to solve the recursion relation $u_{\ell} ^+-  \dfrac{ 2\,\bar{\Omega}}{\bar{q}}
  \left(1+\dfrac{F_0-F_{\ell-1}} {\bar{\Omega}}\right) u_{\ell-1}^+
 + u_{\ell-2}^+ = 0  $ (for $\ell \geq 2 $), supplemented by the initial condition
$ u_1^+=   \bar{\Omega}\,u_0^+ / \bar q $, as obtained from Eq.~\eqref{zero_mode_even}.

We follow the mathematical scheme outlined in Ref.~\cite{Khoo_2019} and define a generating function $ U^+(\rho) $ for the coefficient $\left\{ u_{\ell} ^+\right\}$ such that
\begin{align}
  U^+(\rho)=\sum_{\ell = 0}  ^\infty  u_{\ell} ^+ \, \rho^{\ell} \,,\quad
  u_{\ell} ^+ = \dfrac{1} {\ell !}\, 
 \dfrac{d^\ell \,  U^+(\rho)}  {d  \rho^{\ell} }\Big|_{ \rho =0}.
\label{coeff}
\end{align} 
Using this expression, the recursion relation can be rewritten as
  \begin{align}
 &\sum_{ \ell=2}^\infty 
 \left [ u_{\ell} ^+ - \dfrac{ 2\, \bar{\Omega}}{\bar{q}}
 \left(1+\dfrac{F^R_0} {\bar{\Omega}}\right) 
 u_{\ell-1}^+  + u_{\ell-2}^+\right ] \rho^{\ell} =0
\quad \Rightarrow \quad
U^+(\rho) =  \dfrac{ u_0 \,f_+(\rho)} {\rho_+- \rho_-} 
 \left(\dfrac{1} {\rho_- - \rho}
 -\dfrac{1} {\rho_+ -\rho}\right),\nn
& \rho_{\pm}=s \,\eta \pm \sqrt{s^2 \,\eta^2-1} \,, \quad
s=\dfrac{\bar{\Omega}} {\bar{q}}\,,\quad
\eta=1+\dfrac{F_0^R} {\bar{\Omega}}\,,\quad
f_+(\rho)=1-s \,\rho 
\left(1+ \dfrac{ 2\, F^R_0} {\bar{\Omega}}\right).
\label{sol1}
 \end{align}
Using Eqs.~\eqref{coeff} and \eqref{sol1}, we find that
\begin{align}
\label{eqdisp}
   u_{\ell} ^+ = \dfrac{u_0^+}{ \rho_+- \rho_-}
  \left[\dfrac{f_+( \rho_-)}
  { \rho_-^{\ell+1}  }-\dfrac{f_+( \rho_+)}{ \rho_+^{\ell+1}  }\right].
\end{align}

\begin{figure}[]
\centering
\subfigure[]{\includegraphics[width= 0.45 \textwidth]{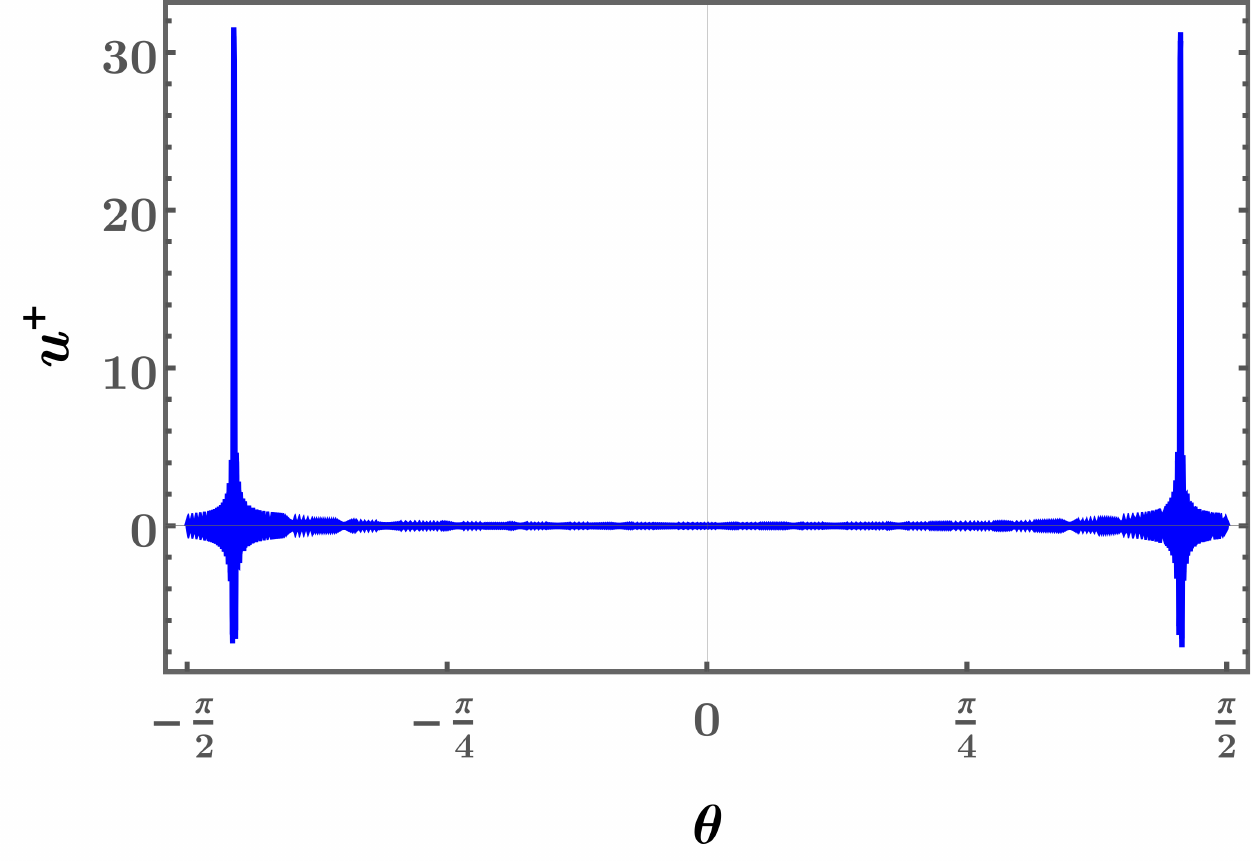}}
\hspace{1 cm}
\subfigure[]{\includegraphics[width= 0.45 \textwidth]{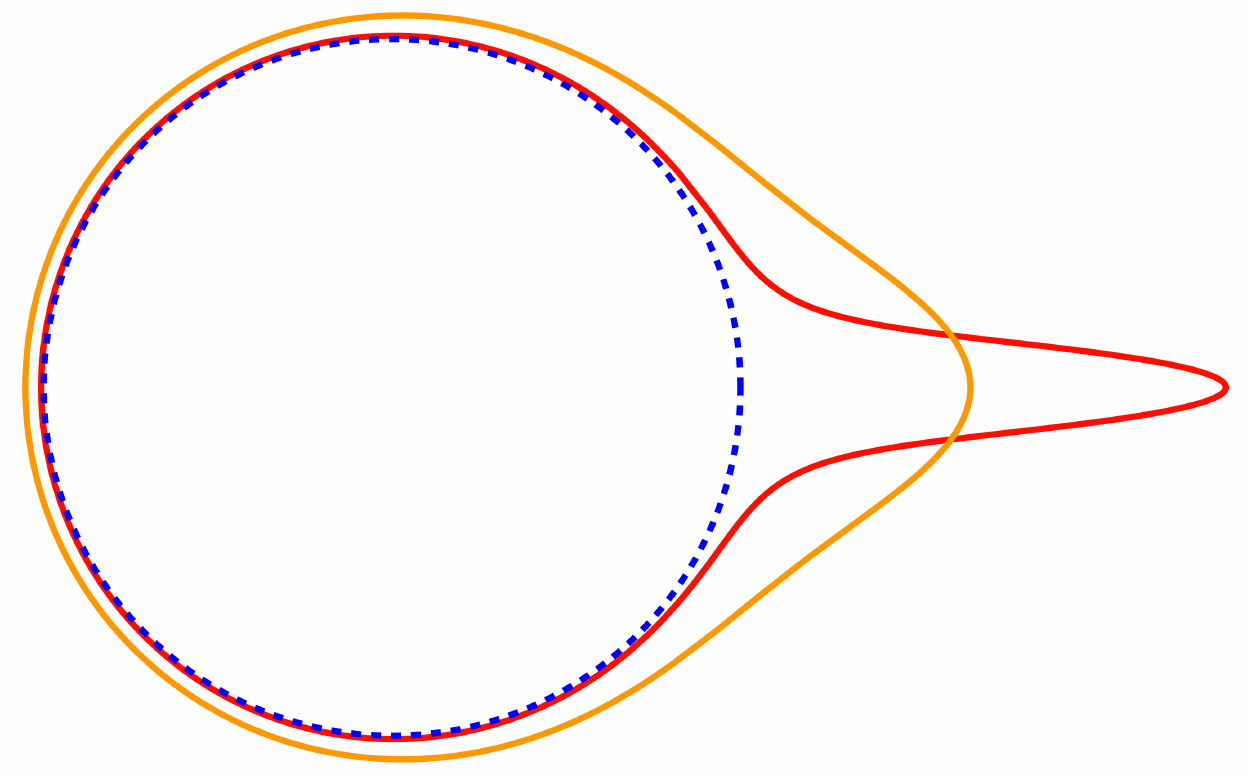}}
\caption{
(a) Behaviour of $u^+$ as a function of $\theta $ for $s\,\eta<1$, obtained from
Eq.~\eqref{eqdisp}. We have used $\bar \Omega =0.01 $, $\bar q = 0.1$, and $\alpha =0.1 $.
The blue narrow peaks show localized excitations belonging to the particle-hole continuum.
(b) The dotted blue circle is the schematic representation of the original (unperturbed) Fermi surface with $k_F=5$. The solid red (for $\alpha=0.01$) and orange (for $\alpha =0.1$) curves represent the perturbed Fermi surface after superimposing the collective mode displacement existing for $s\,\eta >1 $ and $\bar q =0.01$, whose dispersion obeys Eq.~\eqref{zeromod_eq}. 
\label{even_soln}
}
 \end{figure} 

Eq.~\eqref{eqdisp} reveals that there are two kinds of excitations possible depending on whether $s\,\eta<1$ or $ s\,\eta >1$:
\begin{enumerate}

\item
For $s\, \eta<1$, $ \rho_{+}$ is the complex conjugate of $\rho_-$, with $ |\rho_{\pm}|=1$. As a result, the coefficients $ u^+_{\ell} $'s oscillate with $\ell$, and do not diverge.
They give rise to localized solutions for $u^+(\theta)$, which are analogues of the particle-hole excitations of an FL, and their dispersions form an energy band. The boundary for the dispersion of this continuum of excitations is given by the condition $s\, \eta=1 \Rightarrow
  \bar{\Omega}+ F^R_0(a) = \bar q $. In other words, the system supports excitations localized on the Fermi surface for $ \bar{\Omega}+ F^R_0(a) < \bar q $.
We solve this boundary condition for different values of the interaction strength $\alpha$, and plot the resulting curves in Fig.~\ref{ph_zero}(a). 
From Eq.~\eqref{I1final}, we infer that for $a \ll 1$, $ F_0(a)=
\left( 2 \,\bar \Omega \, \alpha \right) ^{2/3}  /\pi 
+ \mathcal{O} \big ( a^{4/3} \big ) $, and for $a>>1$,
$ F^R_0 = {2}/{\pi}$. The physically relevant region of course corresponds to the $\alpha <1 $ region. The boundary condition in this regime can then be approximated by $\bar{\Omega} \left[1 + 
\left( 4  \, \alpha^2 /{\bar \Omega}\right) ^{1/3} \right ] /\pi = \bar q$.
There is a energy crossover scale $  \Omega = { \Omega}^*\approx 
4\,v_F\, k_F \,\alpha^2
= \dfrac{e^4} { 16 \,\pi^2\, v_F\,k_F } $, such that the curve goes as (a) $  \bar \Omega \sim {\bar q}^{3/2}$ for ${  \Omega} < {\Omega}^*$, and (b) $  \bar\Omega \sim \bar q$ for ${\Omega} > {\Omega}^*$. Hence, for very small values of $\alpha$, ${\Omega}^*$ can be quite large, giving us a small range of $\Omega  \in [0, \Omega^*)$ where the behaviour $\propto \bar q^{3/2}$ dominates. This is demonstrated in Fig.~\ref{ph_zero}(b). The main conclusion from this discussion is that for low-enough values of $\Omega$, the behaviour of the boundary curve goes as ${\bar q}^{3/2}$ for our NFL case, unlike the FL case where it goes as linear-in-$\bar q$. The localized nature of these excitations is captured in Fig.~\ref{even_soln}(a).

\item
For $s\, \eta>1$, we have $ \rho_+>1$ and $ \rho_-<1$. Consequently, $ f_+ (\rho_-) / 
{ \rho_-^{\ell+1}} \rightarrow \infty$ as $\ell \rightarrow \infty $, and the first term inside the bracket of the right-hand-side of Eq.~\eqref{eqdisp} blows up unless $f_+ (\rho_-) $ vanishes. Hence, in order to get a physical solution, we must have
\begin{align}
f_+( \rho_-) &= 0  \quad \Rightarrow \quad
\left(\bar{\Omega}+  F^R_0 \right)
-\dfrac{\bar{q}^2} {\bar{\Omega}+2 \,  F^R_0}
=\sqrt{(\bar{\Omega} +   F^R_0)^2-\bar{q}^2}
\nn  & \Rightarrow 
\bar{\Omega} \left(\bar{\Omega}+2 \,   F^R_0\right)
=\bar{q}^2 \,.
 \label{zeromod_eq}
\end{align} 
The dispersion obtained form this relation corresponds to isolated solutions analogous to ``bound states'' created by a bond disorder in a tight-binding Hamiltonian. Here, the source of the bond disorder is $F_0^R$. Hence, in this case, instead of getting a continuum (or energy band), we get discrete energy levels. In this simplified model with $F^R_{\ell>0} = 0$, we have in fact a single isolated bound state that can exist only for $F_0^R > 0$, and it is the analogue of the zero sound mode of the FL. Using this analogy, let us call this NFL zero mode also a zero sound mode.

The dispersion corresponding to Eq.~\eqref{zeromod_eq} is shown in Fig.~\ref{ph_zero}(c) for different values of $\alpha$. For $a \ll 1$, we have
$ \bar{\Omega}^2
\left[  1 +  
\left( \frac{8\times 4  \, \alpha^2 } { \bar \Omega \,\pi^3} \right) ^{1/3}   \right]
\simeq {\bar q}^2  $, which indicates that there exists an energy crossover scale $  \Omega = {\tilde \Omega}^* \approx
 32 \,v_F\, k_F \,\alpha^2 /\pi^3 $, such that the dispersion shows contrasting asymptotes across this value. More explicitly, we find that the dispersion relation reduces to
(a) $ \bar{\Omega} \,
\sqrt{  2 \left( 4  \, \alpha^2 / \bar \Omega \right) ^{1/3}  /\pi}
\simeq {\bar q} 
\Rightarrow 
\bar \Omega \sim {\bar q}^{6/5}  $
for ${\Omega} <  {\tilde \Omega}^* $, and (b) $  \bar\Omega \sim \bar q$ for
${\Omega} >  {\tilde \Omega}^* $. Hence, for very small values of $\alpha$, we have a narrow range of $\Omega  \in [0,  {\tilde \Omega}^*)$ where the behaviour $\propto \bar q^{6/5}$ dominates. This feature is captured in Fig.~\ref{ph_zero}(d). This leads to the crucial conclusion that for low-enough values of $\Omega$, the dispersion of the zero sound goes as ${\bar q}^{6/5}$ for our NFL case, unlike the linear-in-$\bar q$ dispersion of the zero sound of an FL. This agrees with the results obtained in Ref.~\cite{else}, where a collective mode featuring $\Omega \sim |\mathbf q|$ is ruled out. The shape of the Fermi surface in the presence of the zero sound is shown in Fig.~\ref{even_soln}(b).

\end{enumerate}  

\subsection{Odd parity sector}

We now consider the odd parity solutions of Eq.~\eqref{QBE1}, given by $u_\ell^- $ of Eq.~\eqref{eq_decomp}.
We follow the same scheme for calculating $u_\ell^- $ as has been followed in the earlier section. In this case, we define the generating function
 \begin{align}
  U^-(\rho) = \sum_{\ell=1} ^\infty  u_{\ell} ^-  
 \, \rho^\ell \,, \quad
 u^-_\ell
 =\dfrac{1} {\ell!} \dfrac{d^\ell\,  U^-(\rho) } 
 {d   \rho^\ell  }\Big|_ {\rho =0 }  \,,
 \end{align}
With the help of these definitions, the recursion relation in Eq.~\eqref{zero_mode_odd} can be rewritten as
\begin{align}
& \sum_{ \ell=2}^\infty 
 \left [  \bar{\Omega} \, u_{\ell} ^- -\bar{q}\, 
\frac{ u_{\ell+1}^-+ u_{\ell-1}^- }   {2}
  + u_{\ell} ^-\left(F^R_0- F^R_{\ell} \right)
  \right ] \rho^{\ell} =0
\quad \Rightarrow \quad
U^-(\rho) = \dfrac{   u^-_1\, f_-(\rho)} { \rho_+- \rho_-}
 \left(\dfrac{1}{ \rho_- - \rho}-\dfrac{1} { \rho_+-\rho}\right),\nn
& \rho_{\pm} =s \,\eta \pm \sqrt{s^2 \,\eta^2-1} \,, \quad
s=\dfrac{\bar{\Omega}} {\bar{q}}\,,\quad
\eta=1+\dfrac{F_0^R} {\bar{\Omega}}\,,
\quad
u^-_l=\dfrac{u_1^-} { \rho_+- \rho_-}
\left[\dfrac{f_-( \rho_-)}{ \rho_-^{\ell+1}  }
-\dfrac{f_-( \rho_+)} { \rho_+^{\ell+1}  }\right],
\quad f_-(\rho)= 
\rho \left(1- \dfrac{2\, \rho \,  F_1^R } {\bar{q}}\right) .
\label{sol2}
 \end{align}
The initial condition from Eq.~\eqref{zero_mode_odd} translates to $ \left ( \bar{\Omega} + F^R_0- F_1^R 
  \right ) u_1^- =  \bar{q}\, u_2^- / 2 $, which gives us
\begin{align}
u_2^-= \dfrac{ 2\,\bar{\Omega} }{\bar{q}}
\left(1+\dfrac{  F^R_0- F_1^R } {\bar{\Omega}}\right).  
 \end{align}
Just like the even sector, for $s\, \eta<1$, we find a region of localized particle-hole type continuum of excitations, whose boundary dispersion is given by $ s\, \eta=1$. On the other hand, for $s\, \eta>1$, in order to have a normalizable solution, we must have $f_-( \rho_-)=0$. Since for $F^R_{\ell>1} =0$, we have $f_-( \rho_-)= \rho_- \neq 0$ and, therefore, no collective mode can exist in this case. 

Guided by the results above, we now add the effects of a nonzero $F^R_1$ in our simplified model. In this case, we find that the condition for the boundary of particle-hole excitations remains the same just (i.e., $s\,\eta =1$), but the condition for existence of the discrete energy collective mode becomes
 \begin{align}
2\,  \rho_- \,  F_1^R (a)= \bar{q}\,.
\end{align}
We numerically find that the above equation does not have any solution for any $\bar{q}$. We can understand this result intuitively from the fact that $  \rho_- / \bar q\approx {\mathcal O}(1)$, whereas $F^R_1 \ll 1$ [see Fig.~\ref{Landau}(b)], for the typical ranges of $ a $ relevant for the problem. 

Next, we make $F^R_2$ nonzero and after carrying out the similar set of steps, we arrive at the condition
\begin{align}
\bar{q}^2- 2 \,  \rho_- \,  F_1^R\, \bar{q}- 4 \, \rho_-^2 \,F^R_2
\left(\bar{\Omega} +   F^R_0 - F_1^R  \right) = 0 
\end{align}
for the dispersion of a possible collective mode.
Again, we numerically find that the equarion does not have any solution because of the small magnitudes of $ F_1^R $ and $F^R_2$. Adding nonzero $ F^R_{\ell} $'s with $\ell >2$ do not change the situation due to the fact that $ F^R_{\ell >0} \ll 1$. As a result, we conclude that no collective mode exists in the odd parity sector.

\section{Effect of Landau parameters in higher angular momentum channels}
\label{secnum}

\begin{figure}[]
 \centering
\subfigure[]{\includegraphics[width= 0.4 \textwidth]{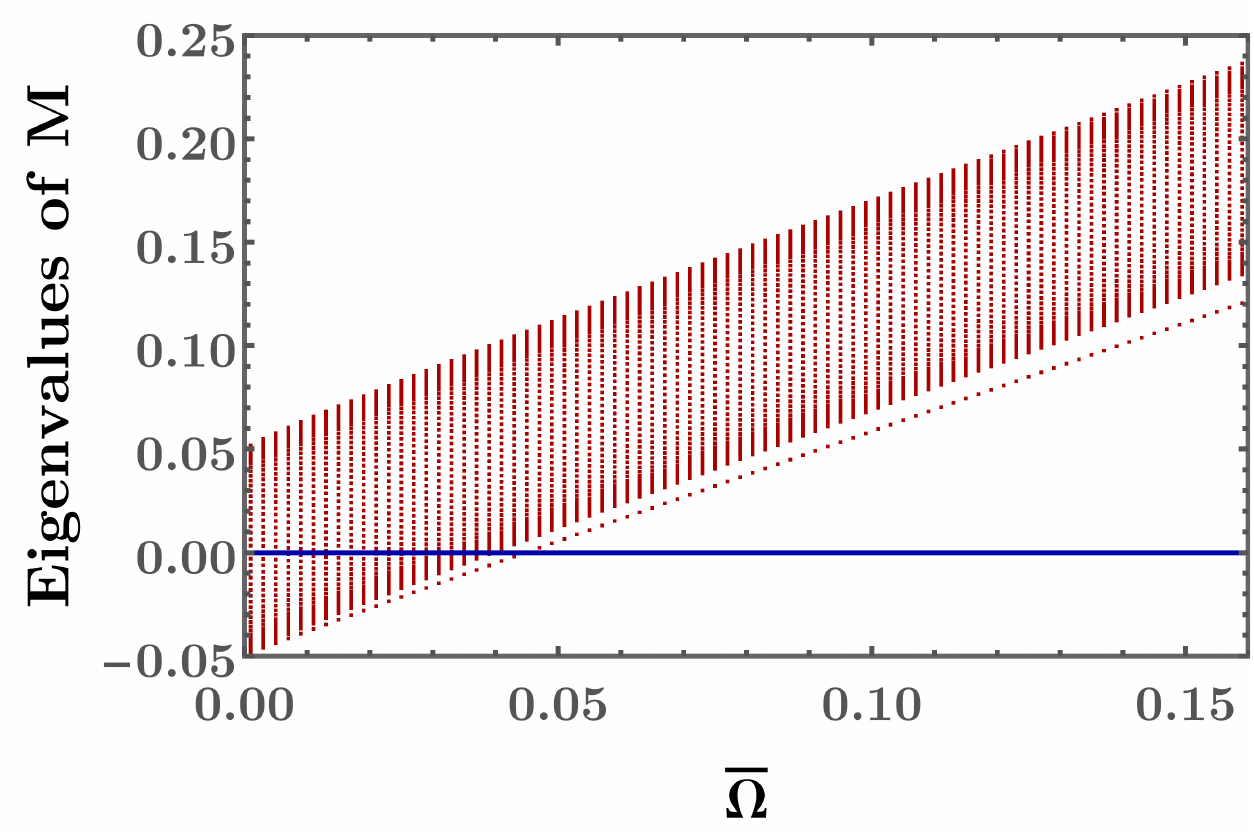}}
\hspace{1 cm}
\subfigure[]{\includegraphics[width= 0.4 \textwidth]{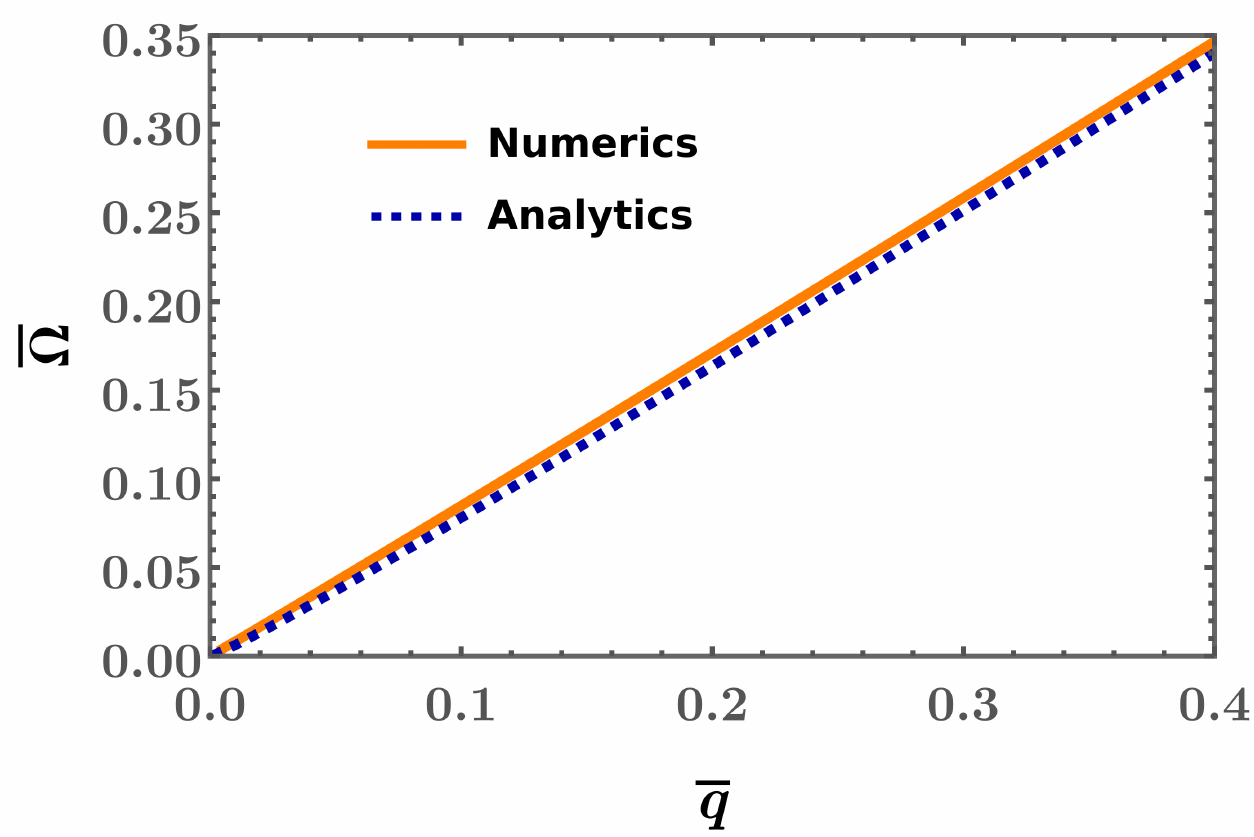}}
\caption{(a) All the eigenvalues of the matrix $M$ [cf. Eq.~\eqref{eq_matrix}], as functions of $\bar{\Omega}$ for $\bar{q} =0.05 $ and $\alpha =0.1$, are shown. The blue line cuts the zero eigenvalues $\lbrace E_0 \rbrace $. The existence of at least one zero eigenvalue ensures the existence of a solution for the Fermi surface displacement $u(\theta)$. Hence, when the blue line cuts the hatched region, we have the particle-hole like excitations. On the other hand, when the blue line intersects the isolated red line, it indicates a collective mode of the Fermi surface.
(b) Dispersion of the collective mode excitation for $\alpha =0.2$, where the orange curve represents the numerical result obtained from diagonalizing $M$, and the dashed blue curve represents our result from the simplified model considered in Sec.~\ref{seceven}.
\label{fig_comp}
}
\end{figure} 

In order to have a more complete calculational framework, here we supplement our simple models in the earlier section with a more realistic scenario, where we include a large number of nonzero $F^R_{\ell} $ components. This model has to be solved numerically, and we show the numerical results for some representative sets of parameter values.

More specifically, for a given value of $n$, we set $F_{\ell> n} =0$, keeping $F_{\ell \leq n} $ nonzero.
Following the same approach as in the earlier section, we now have the recursive equations
\begin{align}
\label{bulk_con}
 u_{\ell} - 2 \,s\,\left(1+\dfrac{  F^R_0} {\bar{\Omega}} \right) 
  u_{\ell-1}+ u_{\ell-2}=0 \text{ for } \ell \geq n+2 \,,
\quad s = \dfrac{\bar{\Omega}} {\bar{q}}\,,
\end{align}
accompanied by the $(n+1)$ boundary conditions
\begin{align}
u_1= s\, u_0  \text{ and }
 u_{\ell + 1} &= 2 \,s
 \left(1+\dfrac{  F^R_0-F^R_{\ell}} {\bar{\Omega}} \right)
 u_{\ell} - u_{\ell-1}
\text{ for } \ell \in [2, n] \,.
 \label{eq_bc}
\end{align}

Defining the generating function
$U(\rho)=
\sum_{\ell = 0}^\infty  u_{\ell} \,  \rho^{\ell} $ as before, Eq.~\eqref{bulk_con} can be expressed as
\begin{align}
\sum_{\ell=n+2}^\infty \left[ u_{\ell} - 2 \,s\,
\left(1+\dfrac{  F^R_0} {\bar{\Omega}}\right)\,  
u_{\ell-1}+ u_{\ell-2}\right ]  \rho^{\ell} = 0 \,,
\end{align}
leading to
\begin{align}
 U(\rho) & =\dfrac{ {\mathcal F} (\rho) } { \left ( \rho_- - \rho \right )  
 \left ( \rho_+ - \rho \right )} \,,
 \quad \rho_{\pm}= s \,\eta \pm \sqrt{s^2 \,\eta^2-1}\,, \quad
\eta = 1+\dfrac{  F^R_0} {\bar{\Omega}} \,,\nn
{\mathcal F} (\rho) & = 
\left [ 1-2\,s\, \rho
\left(1+\dfrac{  F^R_0}{\bar{\Omega}}\right)+ \rho^2 \right ]
\sum \limits_{\ell = 0}^{n-1} u_{\ell} \,  \rho^{\ell}
 +\left [ 1-2\,s\, \rho
\left( 1+\dfrac{  F^R_0}{\bar{\Omega}}\right) \right] 
u_n\, \rho^n + u_{n+1}\, \rho^{n+1}\,.
\end{align}
After some algebra, we get
\begin{align}
 u_{\ell>n+1} =\dfrac{1}
 { \rho_+- \rho_-}
 \left [ \dfrac{ {\mathcal{F}} ( \rho_-)} { \rho_-^{\ell+1}  }
 -\dfrac{ {\mathcal F} ( \rho_+)} { \rho_+^{\ell+1}  }\right ] .
\end{align} 
We observe that the condition for having localized particle-hole type excitations remains the same as before (i.e., $s\, \eta \leq 1$), and hence the boundary curve $s\, \eta =1$ for the continuum (or the energy band) depends only on $  F^R_0$. On the other hand, the existence of the collective modes arising for $s\, \eta \geq 1$ is possible only when $ {\mathcal F}( \rho_-) = 0 $. Since it is difficult to get an analytic solution for this condition, as it involves a polynomial of degree $( n+1) $ in the variable $\rho$, we need to solve it numerically. The condition for the existence of the discrete collective modes is given by $\det  [M(\bar{\Omega},\bar{q})]=0$, where 
\begin{align}
\label{eq_matrix}
M(\bar{\Omega}, \bar{q})=
\begin{pmatrix}
  \bar{\Omega} & -\bar{q} & 0 & 0 & 0 & \cdots \\
 -   \bar q /2  & \bar{\Omega}+  F^R_0-F^R_1 & 
     -   \bar q /2  & 0& 0 & \cdots\\
     0 & -   \bar q /2   & \bar{\Omega}+  F^R_0-F^R_2
  & -   \bar q /2  & 0& \cdots\\
 \vdots & \vdots & \vdots & \vdots & \vdots & \ddots
  \end{pmatrix}_{N\times N} ,
\quad
u^T=\begin{pmatrix}
 u_0 & u_1  & \cdots & u_{N-1} 
 \end{pmatrix} , 
 \end{align} 
with $(N-1)$ denoting the cutoff value of $\ell$ in the set of $\lbrace u_\ell \rbrace$
that we consider.
In our numerics, we have set $n=40$ and $N = 50 $. We plot our results in Fig.~\ref{fig_comp}. Subfigure (a) shows all the eigenvlaues of the matrix $  M(\bar{\Omega},\bar{q})$ as functions of $\bar{\Omega}$ for $\bar{q}=0.05 $ and $\alpha= 0.1$. The existence of at least one zero eigenvalue $E_0$ ensures the existence of a solution for the Fermi surface displacement $u(\theta)$. If we find a range of values of $\bar \Omega $ where $E_0$ exists, then that denotes the particle-hole continuum. On the other hand, if $E_0$ is found at a discrete value of $\Omega$, then that corresponds to a bound state or a collective mode excitation.
In Fig.~\ref{fig_comp}(b), we plot the collective mode dispersion obtained numerically (for $\alpha =0.2 $), and we compare it with our analytical result found from the simplified model in Sec.~\ref{seceven}. We find a very good agreement between the numerical and analytical results.

 \section{Summary and outlook}
 \label{secsum}
  
  In this paper, we have analysed the nature of generic displacements of the critical Fermi surface for the NFL system arising at the Ising-nematic QCP. We have used the quantum Boltzmann equation formalism by incorporating a generalized fermion distribution function applicable for our NFL scenario, and have derived a set of recursive equations whose solutions give us the functional forms of the displacements as well as their dispersions.
The excitations include (a) a continuum of localized particle-hole like excitations and (b) collective modes with discrete energies (isolated from the continuum). Although we find a collective mode analogous to the zero sound of a Fermi liquid, for very low values of $\Omega$, it exhibits a $\Omega \sim |\mathbf q|^{6/5}$ dispersion, unlike the $ \Omega \sim |\mathbf q|$ dependence of the zero sound collective mode of a Fermi liquid. However, this fractional power dependence crosses over to the usual $ \Omega \sim |\mathbf q|$ behaviour as we increase $\Omega $ above a crossover scale $\tilde \Omega^*$.
The curve representing the boundary for the particle-hole continuum in the $\Omega$--$\mathbf q$ plane also displays similar crossover characteristics. In particular, the boundary curve shows a crossover from $\Omega \sim |\mathbf q|^{3/2}$ to $ \Omega \sim |\mathbf q|$ behaviour, determined by another crossover frequency scale $\Omega^*$.
  
It is useful to point out that in Ref.~\cite{ips-zero-mode}, the conclusion was that a zero mode exists for this NFL with a linear-in-momentum dispersion, just like the Fermi liquid case. This resulted from the fact that the frequency-dependence of the generalized Landau parameter $  F_{Landau} ( \theta,\nu ) $
was approximated as
\begin{align}
\frac{ F_{Landau} ( \theta,\nu )  } {e^2\,m} =\begin{cases}
\frac{ 1 } { k_F^2\, \theta_{crit}^{2} } = 
\frac{\pi^{2/3} \,v_F^{2/3}}  { e^{4/3} \,m^{2/3}\,|\nu|^{2/3} } 
& \text{ for } |\theta| < \theta_{crit}\\
\frac{ 1 } { k_F^2\,\theta^2} & \text{ for } |\theta| > \theta_{crit}
\end{cases}\,,
\end{align}
where $ \theta_{crit} \simeq 
\frac{ e^{2/3} \,m^{1/3}\,|\nu|^{1/3} }
{ \pi^{1/3} \,v_F^{1/3} \,k_F} $ is a frequency-dependent cutoff in the angular variable.
Here, $F^R$ is the analogue of $ F_{Landau}$ and, as shown in Eq.~\eqref{I1final}, we have not used an approximation for it as was done in Ref.~\cite{ips-zero-mode}. Consequently, we are able to find the excitations spanning over all $\ell$-values (large or small), whereas the approach followed in Ref.~\cite{ips-zero-mode} forces one to restrict to the displacement modes belonging to the channels with $\ell < \ell_{crit} \simeq 1/\theta_{crit}$. The most crucial outcome of keeping the full frequency dependence of the generalized Landau parameter is our finding that at the lowest frequency values, the collective mode shows a $|\mathbf q|^{6/5}$ dispersion unlike the zero sound collective mode of a Fermi liquid. This agrees with the findings of Ref.~\cite{else}, where the author does not find the so-called zero sound mode.
 
We would like to emphasize that we have performed our computations at $T=0$, and we have assumed that the bosons are in equilibrium. In future, it will be useful to relax these conditions, and also include the effects of the collision integral (denoted by the function $F^I$ in this paper), which we have set equal to zero while finding the solutions for the Fermi surface displacements.

\section*{Acknowledgments}

We are grateful to Patrick A. Lee for insightful discussions.

\appendix

\section{Green's functions in the Keldysh formalism}
\label{app1}

The Keldysh formalism automatically includes different Green’s functions, defined according to the location of the time argument on the contour.
For the fermionic part of the action shown in Eq.~\eqref{kelaction}, the lesser, greater, time-ordered, and anti-time-ordered Green’s functions are given by the expressions
\begin{align}
\label{eq4gf}
 G^< (t_1,\mathbf r_1; t_2,\mathbf r_2) &= -i
\left \langle
 \psi_+  (t_1,\mathbf r_1)\,
   {\psi}_-^\dagger (t_2,\mathbf r_2) \right \rangle, \quad
G^> (t_1,\mathbf r_1; t_2,\mathbf r_2) = -i
\left \langle
 \psi^- (t_1,\mathbf r_1)\,  {\psi}_+^\dagger (t_2,\mathbf r_2) \right \rangle,\nn
 G^{\mathcal{T}} (t_1,\mathbf r_1; t_2,\mathbf r_2) &= -i
\left \langle
  \psi_+  (t_1,\mathbf r_1)\,  {\psi}_+^\dagger (t_2,\mathbf r_2) \right \rangle
= \theta(t_1-t_2)\,G^> (t_1,\mathbf r_1; t_2,\mathbf r_2)
+ \theta(t_2-t_1)\,G^< (t_1,\mathbf r_1; t_2,\mathbf r_2),
\nn \text{and }
G^{ \bar{\mathcal{T}}} (t_1,\mathbf r_1; t_2,\mathbf r_2) 
&= -i \left \langle \psi^-  (t_1,\mathbf r_1)\,  {\psi}_-^\dagger (t_2,\mathbf r_2) \right \rangle
= \theta(t_2-t_1)\,G^> (t_1,\mathbf r_1; t_2,\mathbf r_2)
+ \theta(t_1-t_2)\,G^< (t_1,\mathbf r_1; t_2,\mathbf r_2),
\end{align}
respectively.
Analogously, for the bosons, we have the definitions
\begin{align}
D^< (t_1,\mathbf r_1; t_2,\mathbf r_2) &= -i
\left \langle
\phi^+ (t_1,\mathbf r_1)\, {\phi}^- (t_2,\mathbf r_2) \right \rangle, \quad
D^> (t_1,\mathbf r_1; t_2,\mathbf r_2) = -i
\left \langle
\phi^- (t_1,\mathbf r_1)\,{\phi}^+ (t_2,\mathbf r_2) \right \rangle,\nn
 D^{\mathcal{T}} (t_1,\mathbf r_1; t_2,\mathbf r_2) &= -i
\left \langle
 \phi^+  (t_1,\mathbf r_1)\,{\phi}^+ (t_2,\mathbf r_2) \right \rangle
= \theta(t_1-t_2)\,D^> (t_1,\mathbf r_1; t_2,\mathbf r_2)
+ \theta(t_2-t_1)\,D^< (t_1,\mathbf r_1; t_2,\mathbf r_2),
\nn D^{ \bar{\mathcal{T}}} (t_1,\mathbf r_1; t_2,\mathbf r_2) 
&= -i
\left \langle \phi^-  (t_1,\mathbf r_1)\,{\phi}^- (t_2,\mathbf r_2) \right \rangle
= \theta(t_2-t_1)\,D^> (t_1,\mathbf r_1; t_2,\mathbf r_2)
+ \theta(t_1-t_2)\,D^< (t_1,\mathbf r_1; t_2,\mathbf r_2),
\end{align}
From these sets of Green’s functions, the so-called retarded and advanced Green’s functions
(denoted by the superscripts ``$R$'' and ``$A$'', respectively) are obtained as
\begin{align}
\label{eqdefRA}
G^R(t_1,\mathbf r_1; t_2,\mathbf r_2) &= 
G^{\mathcal{T}} (t_1,\mathbf r_1; t_2,\mathbf r_2) 
- G^< (t_1,\mathbf r_1; t_2,\mathbf r_2)
= G^> (t_1,\mathbf r_1; t_2,\mathbf r_2)
- G^{\bar {\mathcal{T}}} (t_1,\mathbf r_1; t_2,\mathbf r_2) \,,  \nn
G^A (t_1,\mathbf r_1; t_2,\mathbf r_2) &= 
G^{\mathcal{T}} (t_1,\mathbf r_1; t_2,\mathbf r_2) 
- G^> (t_1,\mathbf r_1; t_2,\mathbf r_2)
= G^< (t_1,\mathbf r_1; t_2,\mathbf r_2)
- G^{\bar {\mathcal{T}}} (t_1,\mathbf r_1; t_2,\mathbf r_2) \,.
\end{align}
The definitions also imply the following relations:
\begin{align}
G^R(t_1,\mathbf r_1; t_2,\mathbf r_2)
&=\theta(t_1-t_2) \left[ 
G^>(t_1,\mathbf r_1; t_2,\mathbf r_2)-G^< (t_1,\mathbf r_1; t_2,\mathbf r_2)
\right],\nn
G^A (t_1,\mathbf r_1; t_2,\mathbf r_2)
&= -\theta(t_2-t_1) \left[ 
G^>(t_1,\mathbf r_1; t_2,\mathbf r_2)-G^< (t_1,\mathbf r_1; t_2,\mathbf r_2)
\right] ,\nn
G^R (t_1,\mathbf r_1; t_2,\mathbf r_2)& - G^A (t_1,\mathbf r_1; t_2,\mathbf r_2)
= G^> (t_1,\mathbf r_1; t_2,\mathbf r_2)-G^< (t_1,\mathbf r_1; t_2,\mathbf r_2)
\end{align}
for the fermions; and
\begin{align}
D^R(t_1,\mathbf r_1; t_2,\mathbf r_2)
&=\theta(t_1-t_2) \left[ 
D^>(t_1,\mathbf r_1; t_2,\mathbf r_2)- D^< (t_1,\mathbf r_1; t_2,\mathbf r_2)
\right],\nn
D^A (t_1,\mathbf r_1; t_2,\mathbf r_2)
&= -\theta(t_2-t_1) \left[ 
D^>(t_1,\mathbf r_1; t_2,\mathbf r_2)- D^< (t_1,\mathbf r_1; t_2,\mathbf r_2)
\right] ,\nn
D^R (t_1,\mathbf r_1; t_2,\mathbf r_2)& - D^A (t_1,\mathbf r_1; t_2,\mathbf r_2)
= D^> (t_1,\mathbf r_1; t_2,\mathbf r_2)- D^< (t_1,\mathbf r_1; t_2,\mathbf r_2)
\end{align}
for the bosons.

\bibliography{biblio}

\end{document}